\documentclass[10pt]{article}

\usepackage{graphicx}
\usepackage{float}
\usepackage{amsmath}
\usepackage{amsfonts}
\usepackage{jheppub}

\usepackage{color}
\def\gs{g_s}
\def\cD{\cal D}
\def\gW{g_W}

\def\jb{i_{b}}
\def\jt{i_{t}}
\def\eb{\bar{e}}
\def\qb{\bar{q}}
\def\spaa#1.#2.#3{\langle\mskip-1mu{#1}|#2|{#3}\mskip-1mu\rangle}
\def\spbb#1.#2.#3{[\mskip-1mu{#1}|#2|{#3}\mskip-1mu]}
\def\spa#1.#2{\left\langle#1\,#2\right\rangle}
\def\spb#1.#2{\left[#1\,#2\right]}
\def\spab#1.#2.#3{\left\langle#1|#2|#3\right]}
\def\spba#1.#2.#3{\left[#1|#2|#3\right\rangle}
\def\Wzb{\bar{W}_0}
\def\Pzb{\bar{P}_0}
\def\Pthreeb{\bar{P}_3}
\def\P3b{\bar{P}_3}
\def\Ypb{\bar{Y}_p}
\def\Ypz{{Y}_p(z)}
\def\Ywb{\bar{Y}_w}
\def\Ppb{\bar{P}_+}
\def\Pmb{\bar{P}_-}
\def\Wpb{\bar{W}_+}

\def\li{{\rm Li_2}}
\def\g0{\gamma_0}

\def\bentarrow{\:\raisebox{1.3ex}{\rlap{$\vert$}}\!\rightarrow}                 
\def\dkp#1#2#3#4{
\begin{array}{r c l}
#1 & \rightarrow & #2#3 \\
 & & \phantom{\; #2}\bentarrow #4
\end{array}}                                                                    
                                                                               
\newcommand{\cg}{c_\Gamma}

\newcommand{\beq}{\begin{equation}}
\newcommand{\eeq}{\end{equation}}
\newcommand{\beqn}{\begin{eqnarray}}
\newcommand{\eeqn}{\end{eqnarray}}

\newcommand{\nn}{\nonumber}
\newcommand{\Et}{E_t}
\newcommand{\Pt}{P_t}
\newcommand{\pt}{p_t}
\newcommand{\pb}{p_b}
\newcommand{\pw}{p_W}
\newcommand{\pg}{p_g}
\newcommand\tpW        {{\tilde p}_W}
\newcommand\tpb        {\tilde p_b}

\newcommand{\slsh}{\rlap{$\;\!\!\not$}}     % Feynman slash

\newcommand{\as}{\alpha_S}

\newcommand{\gsim}{\mbox{\raisebox{-0.3ex}{%
\footnotesize $\:\stackrel{>}{\sim}\:$}} }

\newcommand{\ep}{\epsilon}

\newcommand{\om}{\omega}

\title{Top-quark processes at NLO in production and decay}

\author{
    John M. Campbell and R. Keith Ellis
    \\
    Fermilab, Batavia, IL 60510, USA
    \\
    E-mails: 
    {\tt johnmc@fnal.gov}, 
    {\tt ellis@fnal.gov}.}

\preprint{
FERMILAB-PUB-12-078-T}

\abstract{
We describe the implementation of top production and decay processes
in the parton-level Monte Carlo program MCFM. By treating the top
quark as being on-shell, we can factorize the amplitudes for top-pair
production, $s$-channel single-top production, and $t$-channel
single-top production into the product of an amplitude for production
and an amplitude for decay. In this way we can retain all spin
correlations.  Both the production and the decay amplitudes are
calculated consistently at next-to-leading order in $\as$. The full
dependence on the $b$-quark mass is also kept.  Phenomenological
results are presented for various kinematic distributions at the LHC
and for the top quark forward-backward asymmetry at the Tevatron.}


\begin{document}

\maketitle

\section{Introduction} 

The advent of the LHC has brought us data samples of top quarks more than 
an order of magnitude bigger than those observed at the Tevatron. Thus it is 
opportune to re-examine the issue of top quark physics to make 
sure that the theoretical treatment is at a level 
appropriate to confront this enlarged data sample. We have therefore 
decided to review the implementation of top processes in MCFM~\cite{Campbell:1999ah,Campbell:2004ch,Campbell:2010ff,MCFMweb}
with a view to providing the most sophisticated next-to-leading (NLO) treatment possible 
within the context of the top pole approximation.

It is now more than twenty five years since the first calculation of the 
NLO QCD corrections to the production of a pair of top quarks 
at a hadron collider~\cite{Nason:1987xz}. This calculation, and subsequent
NLO computations of single top
production~\cite{Tait:1997fe,Bordes:1994ki,Smith:1996ij,Stelzer:1997ns,Harris:2002md,Campbell:2009ss,Campbell:2009gj},
provided predictions for stable top quarks, without consideration of their
subsequent decay. This issue was addressed in Ref.~\cite{Campbell:2004ch}, where
the decay of the top quark was included for the case of single top production, incorporating
also the effect of NLO QCD effects in the decay.
For the case of top pair
production, predictions including the decay of the top quark are now also
available~\cite{Bernreuther:2004jv,Melnikov:2009dn,Bernreuther:2010ny,Campbell:2010ff}.
Top pair production processes in association with a jet, including the decay,
have been considered in ref.~\cite{Melnikov:2011qx}.
In order to consistently
include corrections in the decay of the top quark it is imperative to also
include the effect of NLO on the top quark width, a result that has been available
for a long time~\cite{Jezabek:1988iv}. 
In addition to these approaches, complete off-shell results for the final
state produced by leptonic top quark pair decays, i.e. $pp \to \nu e^+ b e^- \bar\nu \bar b$,
are also now available at NLO~\cite{Bevilacqua:2010qb}. 

The purpose of this paper is to report in detail on a formalism to incorporate
QCD NLO corrections in both production and decay, in a manner which is
easily extensible to other top-production processes. We shall keep only
the diagrams with a resonant top quark propagator, but provide a
complete NLO treatment of both production and decay stages of the
calculation. In particular we will include NLO corrections in the 
decay of the top quark, including from the hadronic decay of the $W$
boson if appropriate, and retain for the first time the effects of a
non-zero bottom quark mass. The price for including the $b$-mass effects
is very modest in most of the calculation, due to the simple structure of the matrix
elements~\cite{Kleiss:1988xr}. Although the top
quark is strictly on shell, we allow the consistent inclusion of the effects of an off-shell
$W$-boson in top decay.

\begin{figure}
\begin{center}
\includegraphics[angle=270,width=14cm]{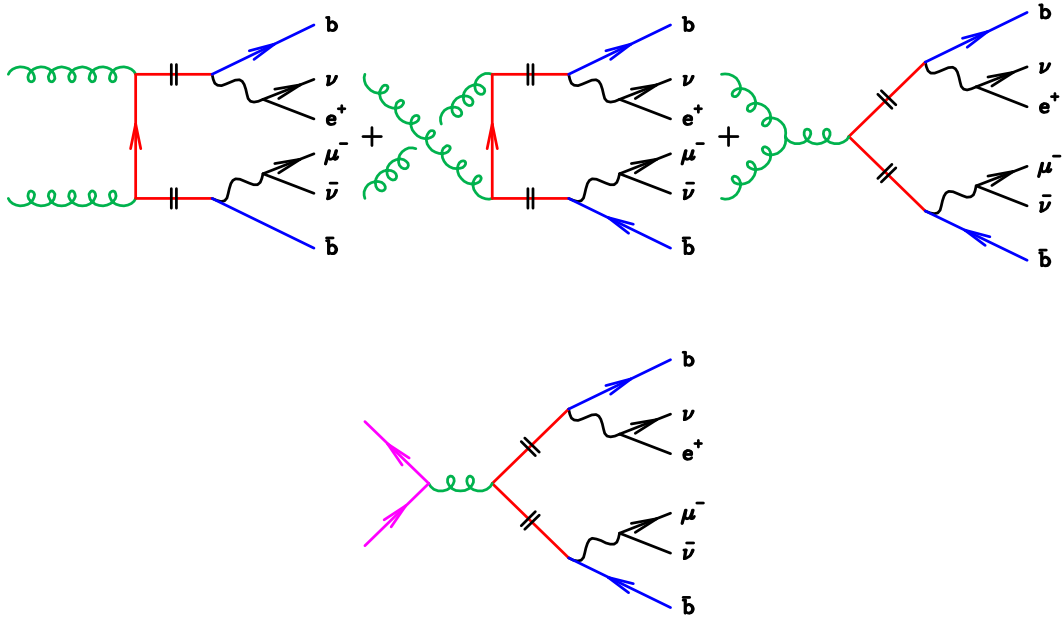}
\caption{Lowest order diagrams for top-pair production and (semi-leptonic) decay.
The double bars indicate that the top (anti-)quark is on shell.}
\label{ttbar}
\end{center}
\end{figure}

\begin{figure}
\begin{center}
\includegraphics[angle=270,width=10cm]{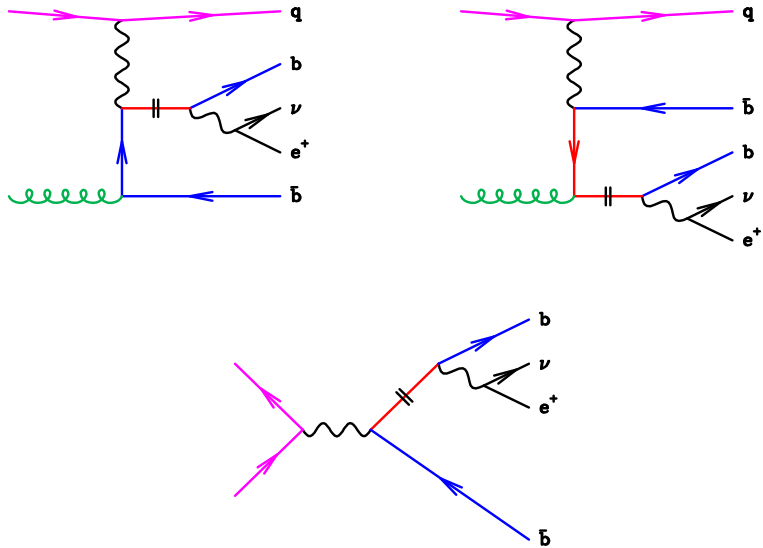}
\caption{Lowest order diagrams for single-top production and (semi-leptonic) decay showing
$t$-channel production (upper line) and $s$-channel production (lower line).
The double bars indicate that the top quark is on shell.}
\label{singletop}
\end{center}
\end{figure}

In this paper we will use this formalism to report on updated results for $t\bar{t}$ pair production, 
single top production with a $W$-boson exchanged in the $t$-channel, and
single top production with a $W$-boson exchanged in the $s$-channel.
However, as indicated above, the formalism should be easily extensible to more
complicated processes such as $t{\bar t}H$, $t{\bar t}Z$ and $t{\bar t}W$.
The lowest order diagrams for top pair production and decay are shown in Figure~\ref{ttbar} and the
lowest order diagrams for single-top production and decay are shown in Figure~\ref{singletop}.
The $s$-channel process has previously been treated in ref.~\cite{Campbell:2004ch},
but now includes the $b$-quark mass consistently throughout.
This paper is the first instance in which 
NLO corrections have been given for the $t$-channel 
single top production process in the four-flavour scheme 
with the decay of the top quark included.
A four-flavour treatment of $t$-channel single top production without 
inclusion of the top quark decay
has previously been given in refs.~\cite{Campbell:2009ss,Campbell:2009gj}.
The top-pair production process has the $b$-quark mass retained throughout,
and in addition includes QCD radiative corrections to the $W$-boson decay
for the case where the top quark decays hadronically. 

The plan of this paper is as follows. In section~\ref{width} we review the results from
the literature on the total width of the top quark, including mass corrections, 
off-shell $W$-boson corrections 
and NLO QCD corrections. Section~\ref{production} illustrates the method which we use to calculate 
top production amplitudes, by reference to the specific case of $s$-channel single top 
production. Section~\ref{decay} presents the results on the amplitudes for top decay, both at the
Born level and the NLO level including both real and virtual corrections.
In section~\ref{sec:realct} we present the counterterm that is used to handle the divergences 
that occur in the real gluon contribution to top decay and
in section~\ref{Radiation_in_decay} we explain our procedure for ensuring that NLO effects
are included consistently in top quark decay.
Section~\ref{pheno} presents an illustration of phenomenological consequences for all three
top production processes. The appendices contain a summary of the spinor notation used throughout
this paper and present a complete calculation of the top quark width at NLO.

\section{Results for the top quark width}
\label{width}
In the limit of very large top mass (dropping the masses of 
the $W$ and the $b,s,d$-quarks), the lowest order result for the top quark width is,
\beq
\Gamma^{(\infty)}=\frac{G_F m_t^3}{8 \pi \sqrt{2}} \sum_{j} |V_{tj}|^2 \, .
\eeq
As usual, $G_F$ is the Fermi constant and $V_{tj}$ are the CKM matrix 
elements for the $t \to j$ transitions.
The superscript $(\infty)$ indicates that this is the result that 
would hold in the limit of infinite top mass, when the masses of the $W$-boson 
and bottom quark can be neglected.
In the following we shall assume $V_{tb}=1,V_{ts}=V_{td}=0$.
In this approximation the finite mass corrections implicit in the 
functions $f$ and $\P3b$ correct this result as follows,
\beq
\Gamma_0(\om^2)=\Gamma^{(\infty)} 2 \P3b f \, .
\label{eq:LOwidth}
\eeq
The notation used in this equation and throughout this paper, 
taken from ref.~\cite{Czarnecki:1990kv}, 
is laid out in Table~\ref{Notation}.
Many of the quantities in Table~\ref{Notation} are functions of the 
variable $z$ which is defined as,
\beq
z =\frac{(p_t-\pw)^2}{m_t^2}\, ,
\eeq
where $p_t$ and $\pw$ are the momenta of the top and the $W$-boson respectively.
For the specific case of the lowest order process we have that $z=\beta^2$, 
corresponding to the $b$-quark being on shell. The variables that depend on $z$,
when evaluated for $z=\beta^2$, are denoted with a bar. Thus we have,
\beq
P_3(\beta^2) \equiv \P3b = \frac{1}{2} \sqrt{\lambda(1,\om^2,\beta^2)}\,,
\eeq
where the function $\lambda$ is also defined in Table~\ref{Notation}.
We note that in the presence of one-gluon emission we have $z_m> z>\beta^2$.
\renewcommand{\baselinestretch}{1.8}
\begin{table}
\begin{center}
\begin{tabular}{|l|l|l|}
\hline
$\beta = \frac{m_b}{m_t}$ & 
$\omega^2 = \frac{\pw^2}{m_t^2}$, &
$z=\frac{(\pt-\pw)^2}{m_t^2}$\\
$\xi=\frac{m_t^2}{m_W^2}$ & 
$\gamma=\frac{\Gamma_W}{m_W}$ & \\ 
$\lambda(x,y,z)=(x-y-z)^2-4 y z $ &
$f=(1-\beta^2)^2+\om^2 (1+\beta^2)-2 \om^4$ &
$z_m=(1-\omega)^2$ \\
$P_0(z)=\frac{1}{2}[1-\om^2+z]$ & 
$P_3(z)=\frac{1}{2}\sqrt{\lambda(1,\om^2,z)}$ & 
$W_0(z)=\frac{1}{2}[1+\om^2-z]$\\
$P_+(z)=P_0+P_3$ & 
$P_-(z)=P_0-P_3$ & 
$Y_p(z)=\frac{1}{2} \ln \frac{P_+}{P_-} $\\
$W_+(z)=W_0+P_3$ & 
$W_-(z)=W_0-P_3$ & 
$Y_w(z)=\frac{1}{2} \ln \frac{W_+}{W_-}$ \\
\hline
\end{tabular}
\caption{Notation used for the calculation throughout this paper.}
\label{Notation}
\end{center}
\end{table}
\renewcommand{\baselinestretch}{1}

The top-quark width is subject to higher order corrections due to gluon radiation, 
\beq \label{Gammawidthexpansion}
\Gamma_t=\Gamma_0(\om^2)+\as \Gamma_1(\om^2) \, .
\eeq
The result for the correction to the width has been given in ref.~\cite{Jezabek:1988iv} 
and in ref.~\cite{Czarnecki:1990kv}\footnote{We follow closely the presentation in 
ref.~\cite{Czarnecki:1990kv}. Eq.~(\ref{totalwidth}) clarifies 
the obvious typographical error $S \to 8$ in Eq.~(27) of ref.~\cite{Czarnecki:1990kv}. 
We thank Andrzej Czarnecki for correspondence on this point. We have slightly modified the form of 
the expression in Eq.~(27) of ref.~\cite{Czarnecki:1990kv} using dilogarithm identities 
in order to make the finiteness in the $\beta \to 0$ limit more manifest.},
\beqn \label{totalwidth}
\as \Gamma_1(\om^2) &=& \Gamma^{(\infty)}  \frac{\as}{2 \pi}C_F \times \nonumber \\
&\Bigg\{ & 8 f \Pzb \Big[\li(1-\Pmb)-\li(1-\Ppb)-2\, \li\left(1-\frac{\Pmb}{\Ppb}\right)
+\Ypb \ln\left(\frac{4 {\P3b}^2}{\Ppb^2 \Wpb}\right)+\Ywb \ln \Ppb\Big] \nonumber \\
&+& 4 (1-\beta^2) \big[ (1-\beta^2)^2+\om^2 (1+\beta^2)-4 \om^4 \big] \Ywb \nonumber \\
&+& [3-\beta^2+11 \beta^4 -\beta^6 +\om^2 (6-12 \beta^2+2 \beta^4)-\om^4(21+5\beta^2)+12\om^6]\Ypb \nonumber \\
&+&8 f \P3b \ln\left(\frac{\om}{4 \P3b^2}\right)+6 \big[ 1-4 \beta^2+3 \beta^4 +\om^2(3+\beta^2)-4\om^4]\P3b \ln \beta \nonumber \\
&+&[5-22\beta^2+5 \beta^4 +9 \om^2 (1+\beta^2)-6 \om^4]\P3b \Bigg\} \, .
\eeqn
Taking the limit $m_b =0$ ($\beta \to 0$) we obtain,
\beqn
&&\as \Gamma_1(\om^2) =  \Gamma^{(\infty)} \frac{\as}{2 \pi} C_F
\Big\{4 (1-\om^2)^2 (1+2 \om^2) \Big[ \li(1-\om^2)-\frac{\pi^2}{3}+\frac{1}{2} \ln(1-\om^2) \ln\om^2\Big] \nonumber \\
&& -2 \om^2 (1+\om^2) (1-2 \om^2) \ln\om^2 
-(1-\om^2)^2 (4 \om^2+5) \ln(1-\om^2)
 +\frac{1}{2} (1-\om^2) (5+9 \om^2-6 \om^4)\Big\} \, .
\eeqn
The effect of the $W$-boson Breit-Wigner propagator on the total width 
can be included as follows~\cite{Jezabek:1988iv},
\beq \label{eq:offshellW}
\Gamma^{BW}_t= 
\frac{\gamma \xi}{\pi} \int_0^{(1-\beta)^2} \, \frac{d \omega^2} {(1-\xi \omega^2)^2+\gamma^2} 
\Big[ \Gamma_0(\omega^2) + \alpha_s \Gamma_1(\omega^2) \Big] \, ,
\eeq
where $\xi$ and $\gamma$ are defined in Table~\ref{Notation}.
The total width for the decay to an on-shell $W$-boson 
can be obtained from Eq.~(\ref{eq:offshellW}) by imposing the narrow width approximation,
\beq
\Gamma^{NW}_t= \Big[ \Gamma_0(\frac{1}{\xi}) + \alpha_s \Gamma_1(\frac{1}{\xi}) \Big] \, .
\eeq
As shown in Table~\ref{widthvalues}
the QCD correction results in a relative change in the top quark width,
\begin{equation}
\frac{\as \Gamma_{1}}{\Gamma_{0}} \approx -0.8 \as
\label{eq:correction_to_width}
\end{equation}
that lowers the leading order result by about 10\%.  
\begin{table}
\begin{center}
\begin{tabular}{|l|c|c|}
\hline
      & $m_W=80.398$~GeV,$m_b=4.7$~GeV   & $m_W=80.398$~GeV,$m_b=0$ \\
\hline
$\Gamma^{BW}_0$ [GeV]     & 1.453518 & 1.457412   \\
$\Gamma^{BW}_1/\Gamma^{BW}_0$  & -0.7878491 & -0.7972087 \\
\hline
$\Gamma^{NW}_0$ [GeV]     & 1.476596   & 1.480522   \\
$\Gamma^{NW}_1/\Gamma^{NW}_0$  & -0.7878090 & -0.7971276 \\
\hline
\end{tabular}
\caption{Values of the LO top width and the NLO correction to the semi-leptonic width,
using $G_F=1.16639\times 10^{-5}$ and $m_t=172.5$~GeV. The results are calculated
using both the full Breit-Wigner for the width (BW) and 
using the narrow width approximation (NW).
In the former case we take $\Gamma_W=2.1054$~GeV. 
The first column represents the complete result for non-zero $W$ and $b$ masses, while
in the subsequent column we approximate by setting the $b$-quark mass to zero.}
\label{widthvalues}
\end{center}
\end{table}

For our purposes we are interested not so much in the total width but rather in 
the pattern of gluon radiation associated with top decay.  Therefore
we need a more differential rate, including both the distributions of
the decay products of the $W$ and of the gluon radiation (if
present). This will be calculated using a point-by-point subtraction
technique that is appropriate for such a calculation~\cite{Ellis:1980wv,Catani:1996vz}.
Thus the integrated real and virtual corrections to the decay rate given in
ref.~\cite{Czarnecki:1990kv} are not sufficient for our purposes.
Nevertheless the computation of the radiative corrections to the 
width contains many essential ingredients for our calculation and also
provides an important input for our result.
Therefore in appendix~\ref{sec:totalwidth} we will present details of the
calculation of the top decay width, in a language that will allow
generalization to the case in which we are interested.

There are further known corrections to the top quark width from 
electroweak effects~\cite{Denner:1990ns,Eilam:1991iz}
and from NNLO QCD~\cite{Czarnecki:1998qc}. 
In this paper we do not require these corrections.
Since we treat the top quark as strictly on shell, the width of the top quark enters 
only as a scale factor, to be chosen to ensure that we get the correct branching ratio
to the appropriate decay channel. We shall comment more on this issue in Section~\ref{Radiation_in_decay}.

\section{Amplitudes for top production}
\label{production}
Our approach relies on a factorization of the calculation into
amplitudes that include the production of a top quark and amplitudes
that represent its subsequent decay. This method is restricted to the
case of top quarks that are produced exactly on shell, but includes
all spin correlations in the decay of the top quark.
By factorizing the calculation in this way we have neglected interference
effects between radiation in the two stages.
Since the characteristic time scale for the production of the top quark is
of order $1/m_t$ while the time for the decay is $1/\Gamma_t$, in general
radiation in the production and decay stages is separated by a large time 
and interference effects are expected to be small, of order
$\as \Gamma_t/m_t$~\cite{Fadin:1993kt,Fadin:1993dz,Melnikov:1993np}.

In this section we will illustrate the method using the case of $s$-channel top production
and briefly describe the implementation of the production amplitudes for all top production
processes that we consider.

\subsection{$s$-channel single top production}
The production of single top by the $s$-channel process is the simplest to
describe. The lowest order process is,
\begin{figure}
\begin{center}
\includegraphics[angle=270,width=5cm]{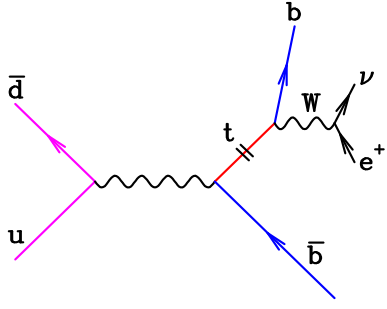}
\caption{Lowest order diagram for $s$-channel single top production.
The double bar indicates that the top is on mass shell.}
\label{schannel-lord}
\end{center}
\end{figure}
\beq
\dkp{u(-u)+\bar{d}(-d)}{\bar{b}(p_c)+}{t(p_t)}{\nu(\nu)+e^+(\eb)+b(p_b)}
\eeq
where the momenta of the particles are shown in brackets in an obvious notation.
We consider all momenta to be outgoing so that,
\beq
u+d+p_c+p_t=u+d+p_c+\nu+\eb+p_b=0.
\eeq
Since our method relies on the assumption that the top quark is exactly on shell, and the bottom quarks are also
on their mass shells we have that, 
\beq
p_t^2=m_t^2,\;\;\;p_c^2=p_b^2=m_b^2.
\eeq
Thus the amplitude for the process in Figure~\ref{schannel-lord} is, 
\beq
M_{\lambda_b \lambda_c} = \delta_{i_d i_u} \delta_{i_b i_c} 
\frac{\gW^2}{2 D_W(s_{ud})} \langle d| \gamma^\rho |u] \; \;
\bar{u}_{\lambda_b}(p_b) \gamma^{\mu} \gamma_L \frac{(\slsh{p}_t+m_t)}{i m_t \Gamma_t} \gamma_\rho \gamma_L v_{\lambda_c}(p_c)
\;\; \frac{\gW^2}{2 D_W(s_{\nu\eb})} \langle \nu| \gamma^\mu |\eb] \;,
\label{eq:fullamplitude}
\eeq
and $\gW^2=4 \sqrt{2} m_W^2 G_F $. 
We have suppressed the colour labels $i_d$, $i_u$, $i_b$ and $i_c$ of the $d$, $u$, $b$ and ${\bar b}$
quarks on the left-hand side. The matrix $\gamma_L$ 
is defined by $\gamma_L = 1-\gamma_R = \frac{1}{2} (1 - \gamma_5)$,
the propagator factor is,
\beq
D_x(s)={(s-m_x^2)+i m_x \Gamma_x} \;,
\eeq
and $s_{ij}=(i+j)^2$, $s_{ijk}=(i+j+k)^2$ where $i$, $j$ and $k$ stand for four-momenta.
The notations $|i\rangle$ and $|j]$ denote spinor solutions of the massless Dirac equation. 
A brief summary of our spinor notation is given in Appendix~\ref{spinnotation}.
Using the identity,
\beq
\sum_{\lambda_t} u_{\lambda_t}(p_t) \bar{u}_{\lambda_t}(p_t) = \slsh{p}_t+m_t \;,
\eeq
we may write Eq.~(\ref{eq:fullamplitude}) as,
\beq
M_{\lambda_b \lambda_c} =\sum_{\lambda_t,i_t} D_{\lambda_b \lambda_t} \, \frac{1}{i m_t \Gamma_t} \, P_{\lambda_t \lambda_c} \;,
\eeq
where the amplitudes representing the production ($P_{\lambda_t \lambda_c}$) and decay ($D_{\lambda_b \lambda_t}$)
of the top quark are given by,  
\beqn
D_{\lambda_b \lambda_t} &=& \delta_{i_b i_t} \bar{u}_{\lambda_b}(p_b) \gamma^{\mu} \gamma_L {u}_{\lambda_t}(p_t)
\;\; \frac{\gW^2}{2 D_W(s_{\nu\eb})} \langle \nu| \gamma^\mu |\eb] \equiv 
\delta_{i_b i_t} \frac{\gW^2}{D_W(s_{\nu\eb})} \bar{u}_{\lambda_b}(p_b) |\nu \rangle [\eb | {u}_{\lambda_t}(p_t) \;,
\label{eq:decayamp} \\
P_{\lambda_t \lambda_c}&=& \delta_{i_t i_c} \delta_{i_d i_u}
\frac{\gW^2}{2 D_W(s_{ud})} \langle d| \gamma^\rho |u] \; \;
\bar{u}_{\lambda_t}(p_t)  \gamma_\rho \gamma_L v_{\lambda_c}(p_c) \nonumber \\ &\equiv&
\delta_{i_t i_c} \delta_{i_d i_u} \frac{\gW^2}{D_W(s_{ud})} \bar{u}_{\lambda_t}(p_t)  |d \rangle [u | v_{\lambda_c}(p_c) \;.
\label{eq:prodamp}
\eeqn
These equations have been simplified on the right hand side by using the Fierz-like identity, Eq.~(\ref{Fierz}).
In order to proceed further we shall use the 
standard trick~\cite{Kleiss:1985yh} of decomposing 
the massive momenta into the sum of two massless momenta, $p_t^\mu = t^\mu + \alpha_t \eta_t^\mu$ with the constant
$\alpha_t$ given by, 
\beq
\alpha_t=\frac{m_t^2}{\langle \eta_t| \slsh{p}_t|\eta_t]} \;.
\eeq
We may write the massive spinors as combinations of massless spinors as follows,
\beqn
u_{-}(p_t)&=&(\slsh{p}_t+m_t) |\eta_t\rangle \frac{1}{\langle t\eta_t\rangle},\;\;\;
u_{+}(p_t)=(\slsh{p}_t+m_t) |\eta_t] \frac{1}{[ t\eta_t]} \;, \\
\bar{u}_{-}(p_b)&=&[\eta_b | (\slsh{p}_b+m_b) \frac{1}{[\eta_b b]},\;\;\;
\bar{u}_{+}(p_b)=\langle \eta_b | (\slsh{p}_b+m_b)  \frac{1}{\langle \eta_b b \rangle } \;, \\
v_{+}(p_c)&=&(\slsh{p}_c-m_b) |\eta_c\rangle \frac{1}{\langle c\eta_c\rangle},\;\;\;
v_{-}(p_c)=(\slsh{p}_c-m_b) |\eta_c] \frac{1}{[ c\eta_c]} \;, \\
\bar{v}_{+}(p_a)&=&[\eta_a | (\slsh{p}_a-m_t) \frac{1}{[\eta_a a]},\;\;\;
\bar{v}_{-}(p_a)=\langle \eta_a | (\slsh{p}_a-m_t) \frac{1}{\langle \eta_a a \rangle} \;.
\eeqn
The momenta $p_t$ and $p_a$ refer to the top and anti-top quark respectively whereas
$p_b$ and $p_c$ refer to the bottom and anti-bottom quark.  
The spin labels of the massless spinors $|\eta_i\rangle,|\eta_i ]$ for $i=t,a,b,c$ 
encode the polarization information 
of the massive quarks and they are equivalent to helicities only in the massless limit.
Stripping the overall colour, coupling and propagator factors,
\beq
P_{\lambda_t\lambda_c} = \delta_{i_t i_c} \delta_{i_d i_u} \, \frac{\gW^2}{D_W(s_{ud})} \, {\cal P}_{\lambda_t\lambda_c} \;,
\eeq
we can evaluate the production amplitudes in Eq.~(\ref{eq:prodamp}) to find,
\beqn
{\cal P}_{--} &=& -
m_b \frac{ \spa{t}.{d} \spb{\eta_c}.{u} }{\spb{\eta_c}.{c}} \to 0\nonumber \\
{\cal P}_{-+} &=& - \spa{t}.{d} \spb{c}.{u}  \nonumber \\
{\cal P}_{+-} &=& -m_b m_t \frac{\spa{\eta_t}.{d} \spb{\eta_c}.{u}}{ \spa{\eta_t}.{t} \spb{\eta_c}.{c}} \to 0
\nonumber \\
{\cal P}_{++} &=& - m_t \frac{\spa{\eta_t}.{d} \spb{c}.{u}}{\spa{\eta_t}.{t}} \to 
m_t \frac{\spa{\eb}.{d} \spb{c}.{u}}{\spa{t}.{\eb}} \;.
\eeqn
The final expressions (in which two of the amplitudes vanish) are valid for the special choices, $\eta_t =\eb, \eta_c= u$.
Unless otherwise stated, we shall always simplify our amplitudes with these choices of $\eta_t$ and $\eta_c$.

\subsubsection{Virtual corrections to $s$-channel single top production.}
\begin{figure}
\begin{center}
\includegraphics[angle=270,width=10cm]{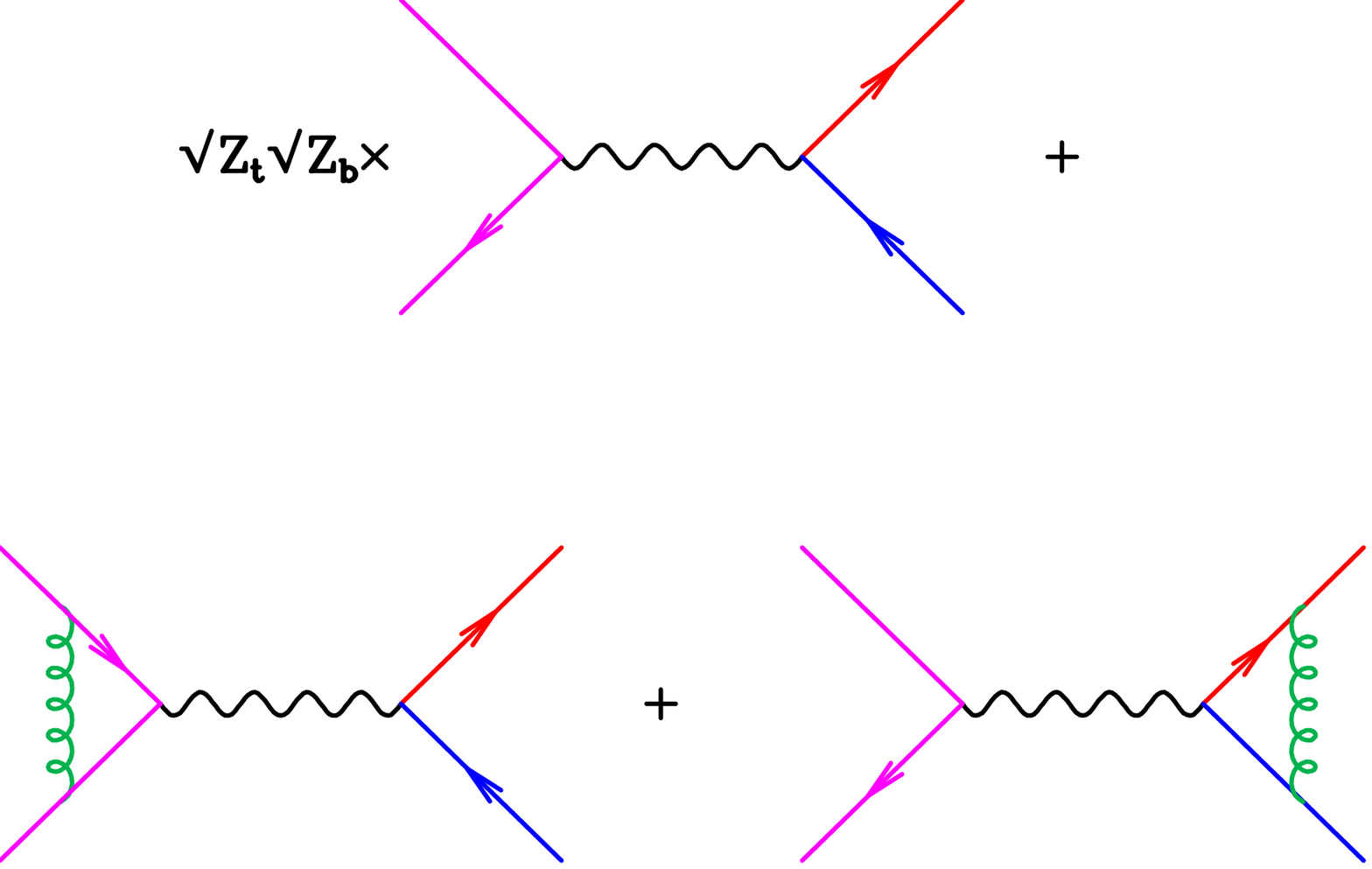}
\caption{Diagram for $s$-channel single top production with virtual gluon radiation.}
\label{schan_virt}
\end{center}
\end{figure}

As is apparent from Figure~\ref{schan_virt} the virtual corrections to the $s$-channel
production are of two types. In the lower row we have the vertex corrections to a timelike
vector boson (a) coupling to two massless fermions and (b) coupling to a massive top quark
and a massive bottom quark. These corrections are well known and will be considered in more detail
in the next section in the context of one-loop corrections to the decay of a top quark.
Indeed the identical corrections occur (a) in the hadronic decay of a W boson, and (b)
in the one-loop corrections to the semileptonic decay of a top quark.
We will therefore defer our discussion of these corrections until later.
\subsubsection{Real corrections to $s$-channel single top production.}
\begin{figure}
\begin{center}
\includegraphics[angle=270,width=10cm]{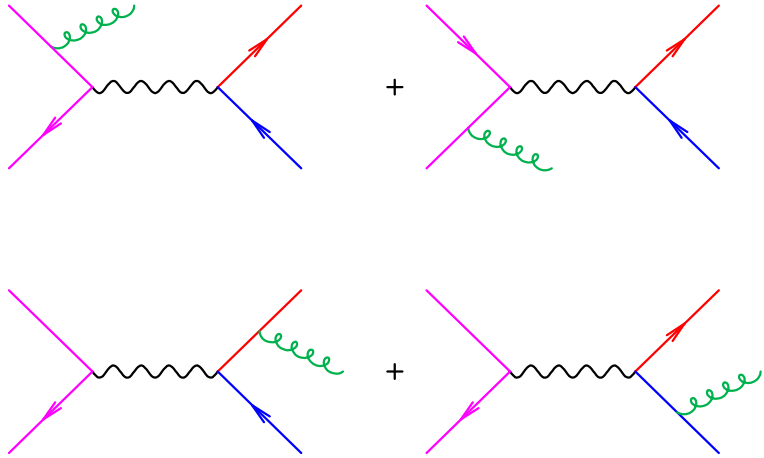}
\caption{Diagram for $s$-channel single top production with additional gluon radiation.}
\label{schan}
\end{center}
\end{figure}

Figure~\ref{schan} shows the diagrams contributing to $s$-channel single
top production in association with an additional radiated gluon. The
diagrams in the top row have gluon radiation from the initial state
quarks, whereas the diagrams in the bottom row have gluon radiation
from the final state quarks.  The two rows of diagrams are separately
gauge invariant and do not interfere because of colour conservation.
In addition to the diagrams in Fig~\ref{schan} there are also diagrams
obtained by crossing the gluon in the top row into the initial state.
The diagrams obtained by crossing the gluon in the bottom 
row into the initial state are not included because they correspond to the 
$t$-channel single-top production process and are therefore accounted for
elsewhere.
We begin by defining the reduced matrix elements for the real radiation 
from the initial and final states,
\beqn
P^{(i)}_{\lambda_t \lambda_g \lambda_c} &=& 
 \gW^2 \gs \left(t^A\right)_{i_d i_u} D_W(s_{udg}) {\cal P}^{(i)}_{\lambda_t \lambda_g \lambda_c} \, , \\
P^{(f)}_{\lambda_t \lambda_g \lambda_c} &=& 
 \gW^2 \gs \left(t^A\right)_{i_t i_c} D_W(s_{ud}) {\cal P}^{(f)}_{\lambda_t \lambda_g \lambda_c} \, ,
\eeqn
where the colour matrices $t^A$ are normalized such that $t^A t^B= \delta^{AB}$.
The spin labels $\lambda_t,\lambda_h,\lambda_c$ encode the polarization state
of the top quark, gluon and bottom anti-quark respectively. 
After a simple calculation we find for radiation from the initial state,
\beqn
   {\cal P}^{(i)}_{---} &=& 0\nonumber \\
   {\cal P}^{(i)}_{--+} &=& \frac{\spab{t}.{d+g}.{u} \spb{c}.{u}}{\spb{g}.{u} \spb{g}.{d}}
          \nonumber \\
   {\cal P}^{(i)}_{+--} &=& 0\nonumber \\
   {\cal P}^{(i)}_{+-+} &=&
       - \frac{m_t\spab{\eb}.{d+g}.{u} \spb{c}.{u}}{\spa{t}.{\eb} \spb{g}.{u} \spb{g}.{d}} 
          \nonumber \\
   {\cal P}^{(i)}_{-+-} &=&\frac{m_b \spa{t}.{d} \spb{g}.{u}}{\spa{g}.{u} \spb{c}.{u}} 
          \nonumber \\
   {\cal P}^{(i)}_{-++} &=&\frac{\spab{d}.{u+g}.{c} \spa{t}.{d}}{ \spa{g}.{u} \spa{g}.{d}}
          \nonumber \\
   {\cal P}^{(i)}_{++-} &=&
       - \frac{m_t m_b \spa{\eb}.{d} \spb{g}.{u}} {\spa{t}.{\eb} \spa{g}.{u} \spb{c}.{u}} 
          \nonumber \\
   {\cal P}^{(i)}_{+++} &=&
       - \frac{ m_t \spab{d}.{u+g}.{c} \spa{\eb}.{d}}{\spa{t}.{\eb} \spa{g}.{u} \spa{g}.{d}} \, .
\eeqn
The results for radiation from the final-state quarks are,
\beqn
   {\cal P}^{(f)}_{---} &=& 0\nonumber \\
   {\cal P}^{(f)}_{--+} &=&
       - \frac{\spa{t}.{d} \spb{c}.{u}}{\spb{g}.{d}} 
\left( \frac{\spab{g}.{p_t}.{d}}{\spab{g}.{p_t}.{g}}
      -\frac{\spab{g}.{p_c}.{d}}{\spab{g}.{p_c}.{g}}\right)
       + \frac{\spa{t}.{g} \spa{g}.{d} \spb{c}.{u}}{\spab{g}.{p_t}.{g}}  \nonumber \\
   {\cal P}^{(f)}_{+--} &=& 0\nonumber \\
   {\cal P}^{(f)}_{+-+} &=&
        \frac{m_t \spa{\eb}.{d} \spb{c}.{u}}{\spa{t}.{\eb} \spb{g}.{d}} 
 \left( \frac{\spab{g}.{p_t}.{d}}{\spab{g}.{p_t}.{g}}
       -\frac{\spab{g}.{p_c}.{d}}{\spab{g}.{p_c}.{g}}\right) 
       + \frac{m_t \spa{g}.{\eb} \spa{g}.{d} \spb{c}.{u} }{\spa{t}.{\eb} \spab{g}.{p_t}.{g}}
          \nonumber \\
   {\cal P}^{(f)}_{-+-} &=&
        \frac{m_b \spa{t}.{d} \spb{g}.{u}^2}{\spb{c}.{u} \spab{g}.{p_c}.{g}}
          \nonumber \\
   {\cal P}^{(f)}_{-++} &=&
           \frac{\spa{t}.{d} \spb{c}.{u}}{\spa{g}.{u}}
\left( \frac{\spab{u}.{p_t}.{g}}{\spab{g}.{p_t}.{g}}
      -\frac{\spab{u}.{p_c}.{g}}{\spab{g}.{p_c}.{g}}\right)
       + \frac{\spa{t}.{d} \spb{c}.{g} \spb{g}.{u}}{\spab{g}.{p_c}.{g}} \nonumber \\
   {\cal P}^{(f)}_{++-} &=&
       - \frac{m_t m_b \spa{\eb}.{d} \spb{g}.{u}^2}{\spa{t}.{\eb} \spb{c}.{u}  \spab{g}.{p_c}.{g}} 
     \nonumber \\
   {\cal P}^{(f)}_{+++} &=&
          - \frac{ m_t \spa{\eb}.{d} \spb{c}.{u}}{\spa{t}.{\eb} \spa{g}.{u}} 
\left( \frac{\spab{u}.{p_t}.{g}}{\spab{g}.{p_t}.{g}}
      -\frac{\spab{u}.{p_c}.{g}}{\spab{g}.{p_c}.{g}}\right)
       - \frac{m_t \spa{\eb}.{d} \spb{c}.{g} \spb{g}.{u}}{\spa{t}.{\eb} \spab{g}.{p_c}.{g}} \, .
\eeqn
\subsection{Top pair production}
In our current treatment of top pair production we have 
followed the procedure outlined above and expressed the complete amplitude as a product of
the amplitude for $t \bar{t}$ production times the amplitude for the decay of the top quark
and the anti-quark. In addition we have coded the analytic results for the one-loop amplitudes
for top pair production given in the paper of Badger et al., ref.~\cite{Badger:2011yu}. These
amplitudes are expressed simply in terms of spinor products. Our previous treatment of the one-loop
amplitudes~\cite{Campbell:2010ff} was based on the work of K{\"o}rner et al.~\cite{Korner:2002hy} 
in which the results
are expressed in terms of more complicated spinor strings.
We note that the two calculations are in complete numerical agreement.
We find that using the new expressions of
Ref.~\cite{Badger:2011yu} in MCFM improves the speed of the 
calculation of the virtual corrections by a factor of three.

\subsection{$t$-channel single top production}
For the $t$-channel single top process we use the four-flavour scheme in which the lowest
order process is,
\beq
q + g \to q^\prime + {\bar b} + t (\to \nu e^+ b) \;,
\eeq
as indicated in the top row of Figure~\ref{singletop}. The next-to-leading order corrections to this process,
without including the top decay products (i.e. $qg \to q^\prime {\bar b} t $), were presented
in Refs.~\cite{Campbell:2009ss,Campbell:2009gj}. In this paper we re-use the amplitudes computed
for that calculation but extend them to incorporate the top quark decay in a fashion similar
to the approach presented for the other two cases above.
A slight difference is that the virtual amplitudes used in Ref.~\cite{Campbell:2009ss,Campbell:2009gj}
had already been simplified by making a
specific choice for the vectors that are used to decompose the massive top quark
momentum, $\eta_t$. In that case the standard decay amplitudes that are
presented in the following section cannot be used. Rather than repeat the calculation of the
relevant amplitudes with the canonical choice of this vector used throughout the rest of this paper,
we have simply used expressions for the Born-level decay amplitudes (see Eq.~(\ref{treeleveltopdecay}) in the next
section) that are appropriate for this alternative choice of $\eta_t$.

\section{Amplitudes for top decay}
\label{decay}
In this section we list the various decay amplitudes that are used in the calculations presented
in this paper. Except where otherwise noted, we make the choices $\eta_t =\eb,\eta_b =\nu$, with
the choice of $\eta_t$ required in order to match the production amplitudes.

\subsection{Top decay at Born level}
The Born decay amplitudes for the process
\beq
t(p_t) \to b(p_b) + \nu(\nu)+e^+(\eb)
\eeq
can be easily evaluated from the expression given in Eq.~(\ref{eq:decayamp}).
In order to report the results using spinor products it is useful
to define the massless vectors, $t\cdot t=b\cdot b=0$, using the auxiliary vectors, $\eta_b,\eta_t$,
\beqn
t^\mu &=& p_t^\mu - \frac{m_t^2}{2 p_t. \eta_t} \eta_t^\mu\, , \nonumber \\
b^\mu &=& p_b^\mu - \frac{m_b^2}{2 p_b. \eta_b} \eta_b^\mu \, .
\label{eq:tandbdefn}
\eeqn
Separating out the colour, coupling and propagator factors,
\beq
D_{\lambda_b \lambda_t }= 
\delta_{\jb \jt}  \, \frac{\gW^2}{D_W(s_{\nu\eb})} \,{\cD}_{\lambda_b \lambda_t} \;,
\eeq
we find,
\beqn \label{treeleveltopdecay}
{\cD}_{--} &=& 
\langle b \nu \rangle [\eb t] \nonumber \\
{\cD}_{-+} &=& 
m_t \langle b \nu  \rangle \frac{[\eb \eta_t]}{[t \eta_t]}  \to 0 \nonumber \\
{\cD}_{+-} &=&  
m_b \frac{\langle \eta_b \nu \rangle }{\langle \eta_b b\rangle }  [\eb t]  \to 0 \nonumber \\
{\cD}_{++} &=& 
m_b m_t \frac{\langle \eta_b \nu\rangle }{\langle \eta_b b\rangle } 
\frac{[\eb \eta_t]}{[t \eta_t]}  \to 0 \, .
\eeqn
Making the choices $\eta_t =\eb$ (as in the production amplitudes) and $\eta_b =\nu$, all but the amplitude ${\cD}_{--}$ vanish.

As noted above, for the particular case of virtual and real radiation corrections to the $t$-channel production process the
amplitudes have already been simplified using $\eta_t=g$, the momentum of the gluon in the initial
state. One can read off the appropriate decay amplitudes from Eq.~(\ref{treeleveltopdecay}) by making the choices
$\eta_t =g,\eta_b =\nu$. We see that in this case the amplitude ${\cD}_{-+}$ does not vanish.

The amplitudes for the related charge-conjugate process,
\beq
{\bar t}(p_a) \to {\bar b}(p_c) + e^-(e)+ \bar{\nu}(\bar{\nu}) \;,
\eeq
can be obtained from the above amplitudes by symmetry. Denoting the colour-stripped amplitudes for this
process by $\overline{\cal D}_{\lambda_a \lambda_c}$ and introducing the operator ${\cal C}$ defined by,
\beq
{\cal C}: \qquad t \to a \;, \quad b \to c \;, \quad \nu \to \bar\nu \;, \quad \bar{e} \to e \;,
 \quad \spa i.j \leftrightarrow \spb i.j \;,
\label{eq:Cdefn}
\eeq
then we have,
\beq
\overline{\cal D}_{\lambda_a \lambda_c} = - {\cal C} \left[ {\cal D}_{-\lambda_b \, -\lambda_t}\right] \;.
\label{eq:DbarDrelation}
\eeq
Note that the massless vectors $a$ and $c$ that appear in the expressions for the amplitudes
are related to the massive four-momenta of the top and bottom quarks, $p_a$ and $p_c$, by the
transformed equivalents of Eq.~(\ref{eq:tandbdefn}),
\beq
a^\mu = p_a^\mu - \frac{m_t^2}{2 p_a. \eta_a} \eta_a^\mu\, , \qquad
c^\mu = p_c^\mu - \frac{m_b^2}{2 p_c. \eta_c} \eta_c^\mu \, ,
\label{eq:aandcdefn}
\eeq
with the choices $\eta_a = e$ and $\eta_c = \bar\nu$.

\subsection{Virtual corrections to top decay}
\label{sec:topdecayvirt}
\begin{figure}
\begin{center}
\includegraphics[angle=270,width=14cm]{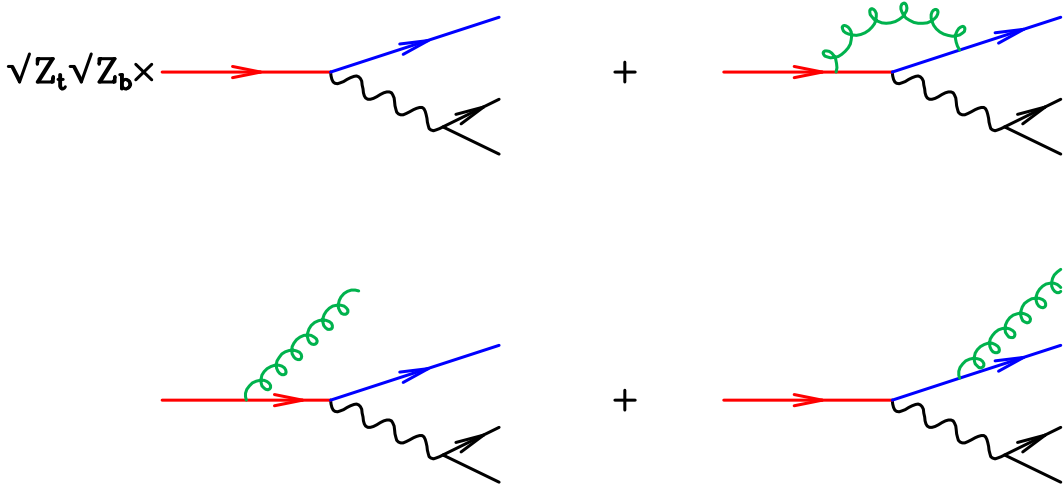}
\caption{Diagrams for gluon radiation from the $t,b$-line in top decay. $Z_b$ and $Z_t$ denote the wave function
renormalization constants for the external fermion lines. The expression for the $Z_q, (q=b,t)$ is given 
in the appendix, Eq.~(\ref{wavefunction}).}
\label{tdecayg}
\end{center}
\end{figure}

We now discuss top quark decay with virtual gluon radiation from the $t,b$ line as shown in 
the top row of Figure~\ref{tdecayg}.
We give the results for the special choices, $\eta_t =\eb, \eta_b= \nu$ for the one loop
corrections to the decay, ${\cD}_{\lambda_b \lambda_t}^V$. The labels $\lambda_b$ and $\lambda_t$ denote the spin state of
the bottom and top quark respectively. In the massless limit they would correspond to helicities.
The general result for the renormalized one-loop form factor in the case $m_t \neq 0, m_b \neq 0$ is,
\beq
{\cal F}_{\mu} (\pb,\pt)=\frac{\gW}{\sqrt{2}} \bar{u}(\pb) \Gamma_{\mu}(\pb,\pt) u(\pt) \;,
\eeq
where $\Gamma_{\mu}(\pb,\pt)$ is decomposed as,
\beqn
\Gamma^\mu(\pb,\pt)&=&\gamma^\mu \gamma_L  
 +g^2 \; C_F \cg 
\Big(\frac{\mu^2}{m_t^2}\Big)^\ep 
\times\nonumber \\ &\Biggl[& 
 C_0^L  \gamma^\mu \gamma_L 
+C_0^R  \gamma^\mu \gamma_R 
+C_1^R \frac{\pb^\mu}{m_t} \gamma_R 
+C_1^L \frac{\pb^\mu}{m_t} \gamma_L 
+C_2^R \frac{\pw^\mu}{m_t} \gamma_R
+C_2^L \frac{\pw^\mu}{m_t} \gamma_L\Biggr]
 +O(\ep) \;. 
\label{eq:massiveF}
\eeqn
The factor $\cg$ is the standard prefactor that appears in
dimensionally regulated one-loop calculations,
\begin{equation} \label{cGamma}
\cg = \frac{1}{(4\pi)^{2-\epsilon}}
\frac{\Gamma(1+\epsilon)\Gamma^2(1-\epsilon)}{\Gamma(1-2\epsilon)} = \frac{1}{(4\pi)^{2-\epsilon}}
\frac{1}{\Gamma(1-\epsilon)}+O(\epsilon^3) \; .
\end{equation}
If we ignore the masses of the decay products of the $W$-boson, the $C_2^{L/R}$ terms 
will not contribute to physical amplitudes. They will be dropped in the 
following. The results for the remaining coefficients are given in 
Eq.~(\ref{resultsforCcoefficients}) of Appendix~\ref{Virtual_corrections}.
The amplitudes $C_0^R,C_1^L$ and $C_1^R$ are finite in the limit $\ep \to 0$.
Removing the common overall factor,
\beq
D^V_{\lambda_b \lambda_t }= \delta_{\jb \jt}\,
\frac{\gW^2 \gs^2 \cg C_F}{D_W(s_{\nu \eb})} \Big(\frac{\mu^2}{m_t^2}\Big)^\ep \, {\cD}^V_{\lambda_b \lambda_t } \;,
\eeq
we have that,
\beqn
{\cD}_{--}^V   &=& 
\Bigg\{ C_0^L   \spa{b}.{\nu} \spb{\eb}.{t} 
+   \frac{C_1^R}{2}  \frac{\spa{b}.{\nu} \spa{\eb}.{b} \spb{\eb}.{b} }{\spa{\eb}.{t}} 
      + m_t m_b  C_0^R  \frac{\spa{\eb}.{\nu} \spb{\nu}.{\eb} }{\spa{\eb}.{t} \spb{\nu}.{b}} 
+ \frac{m_b}{m_t} \frac{C_1^L}{2} 
        \frac{\spa{b}.{\nu} \spb{\nu}.{t} \spb{\eb}.{b}}{\spb{\nu}.{b}}  \Bigg\}
\nonumber \\
{\cD}_{-+}^V &=&  \Bigg\{ 
      m_b C_0^R \frac{\spa{\nu}.{t} \spb{\nu}.{\eb}}{\spb{\nu}.{b}}
        - \frac{C_1^R}{2 m_t}  \spa{\nu}.{b} \spa{b}.{t} \spb{\eb}.{b} 
     + m_b \frac{C_1^L}{2} \frac{\spa{\nu}.{b} \spb{\nu}.{\eb} \spb{\eb}.{b}}{\spb{\nu}.{b} \spb{\eb}.{t}} 
\Bigg\}
\nonumber \\
{\cD}_{+-}^V &=&  \Bigg\{ 
m_t C_0^R  \frac{\spa{\nu}.{\eb} \spb{\eb}.{b}}{\spa{\eb}.{t}} 
-\frac{C_1^L}{2 m_t} \spa{\nu}.{b} \spb{\eb}.{b} \spb{b}.{t} 
      + m_b \frac{C_1^R}{2} \frac{\spa{\nu}.{\eb} \spb{\eb}.{b}}{\spa{\eb}.{t}} \Bigg\}
\nonumber \\
{\cD}_{++}^V&=&  
  \Bigg\{ 
           C_0^R  \spa{t}.{\nu} \spb{\eb}.{b} 
            +\frac{C_1^L}{2} \frac{\spa{b}.{\nu} \spb{\eb}.{b}^2}{\spb{\eb}.{t}} 
           +\frac{C_1^R}{2}\frac{m_b}{m_t}  \spa{t}.{\nu} \spb{\eb}.{b} 
\Bigg\} \;.
\eeqn
Because of the vanishing results in Eq.~(\ref{treeleveltopdecay}), ${\cD}_{+\lambda_t}=0$, 
for our particular choice of auxiliary vectors, 
$\eta_t,\eta_b$ the one-loop results for ${\cD}_{+-}^V,{\cD}_{++}^V$ will not be needed. 

The virtual amplitudes for the decay of an anti-top quark are related to those for the
decay of a top quark in the same manner as for the lowest order amplitudes. Namely,
\beq
\overline{\cal D}^V_{\lambda_a \lambda_c} = - {\cal C} \left[ {\cal D}^V_{-\lambda_b \, -\lambda_t}\right] \;,
\label{eq:DbarVDVrelation}
\eeq
where the operation ${\cal C}$ is defined in Eq.~(\ref{eq:Cdefn}).

\subsection{Top decay with gluon radiation}
We now discuss top quark decay with gluon radiation from the $t,b$ line as shown in 
the bottom row of Figure~\ref{tdecayg}.
In order to report the results using spinor products it is useful
to define the massless vectors, $t\cdot t=b\cdot b=0$,  
as in Eq.~(\ref{eq:tandbdefn}), 
using the auxiliary vectors, $\eb,\nu$,
\beq
t^\mu = p_t^\mu - \frac{m_t^2}{2 p_t. \eb} \eb^\mu\, , \qquad
b^\mu = p_b^\mu - \frac{m_b^2}{2 p_b. \nu} \nu^\mu \, .
\eeq
We further define the colour-stripped decay amplitudes through the relation,
\beq
D^R_{\lambda_b \lambda_g \lambda_t }= 
\gs \mu^\ep  (t^A)_{\jb \jt}\, \frac{\gW^2}{D_W(s_{\nu\eb})} \,  {\cD}^R_{\lambda_b \lambda_g \lambda_t } \;,
\eeq
where the colour matrices $t^A$ are normalized such that $t^A t^B= \delta^{AB}$.
The explicit expressions for the decay amplitudes are,
\beqn
      {\cD}^R_{---}&=& 
     \frac{\spa{b}.{\nu}\spb{t}.{\eb}}{\spb{b}.{g}}   
     \Bigg(\frac{\spab{g}.{p_b}.{b}}{\spab{g}.{p_b}.{g}}
         -\frac{\spab{g}.{p_t}.{b}}{\spab{g}.{p_t}.{g}}\Bigg)
      +\frac{\spa{b}.{g} \spa{g}.{\nu} \spb{t}.{\eb} }{\spab{g}.{p_b}.{g}} \;,
\nonumber \\
      {\cD}^R_{-+-}&=&   
     \frac{\spa{b}.{\nu}\spb{t}.{\eb}}{\spa{g}.{b}}   
     \Bigg(\frac{\spba{g}.{p_b}.{b}}{\spba{g}.{p_b}.{g}}
         -\frac{\spba{g}.{p_t}.{b}}{\spba{g}.{p_t}.{g}}\Bigg)
      +\frac{\spa{b}.{\nu}  \spb{g}.{t} \spb{g}.{\eb}}{\spba{g}.{p_t}.{g}}   \;, 
\nonumber \\
      {\cD}^R_{+--}&=& \frac{m_b\spa{\nu}.{g}^2 \spb{\eb}.{t} }
       {\spa{\nu}.{b} \spab{g}.{p_b}.{g}}  \;,
\nonumber \\
      {\cD}^R_{++-}&=& 0  \;,
\nonumber \\
      {\cD}^R_{--+}&=& 0  \;,
\nonumber \\
      {\cD}^R_{-++}&=&  \frac{m_t\spa{\nu}.{b} \spb{\eb}.{g}^2 }
      {\spb{\eb}.{t} \spab{g}.{p_t}.{g}}   \;, 
\nonumber \\
      {\cD}^R_{+-+}&=& 0  \;,
\nonumber \\
      {\cD}^R_{+++}&=& 0  \;.
\eeqn
As expected the soft gluon singularities are all contained in the eikonal terms
for ${\cD}^R_{---}$ and ${\cD}^R_{-+-}$. These can be simplified using the Schouten identity,
Eq.(\ref{Schouten}), to remove the unphysical singularity. 
For example, the amplitude $ {\cD}_{---}$ contains the singular term,
\beq
     \frac{1}{\spb{b}.{g}}   
     \left(\frac{\spab{g}.{p_b}.{b}}{\spab{g}.{p_b}.{g}}
          -\frac{\spab{g}.{p_t}.{b}}{\spab{g}.{p_t}.{g}}\right) = 
\frac{\spaa{g}.{\slsh{p}_b \slsh{p}_t}.{g}}{\spab{g}.{p_b}.{g} \spab{g}.{p_t}.{g}} \;.
\eeq
Squaring the part of the full decay amplitude that contains 
the eikonal factor we obtain the standard expression,
\beq
\left|D^R_{---}\right|^2=\gs^2 \mu^{2 \epsilon} 
Tr (t^A t^A) \left|\frac{\spaa{g}.{\slsh{p}_b \slsh{p}_t}.{g}}{\spab{g}.{p_b}.{g} \spab{g}.{p_t}.{g}}\right|^2=
\frac{1}{2} \gs^2 \mu^{2 \epsilon} C_F \times \Big[ -\frac{m_b^2}{(p_b.g)^2} +\frac{2 p_t . p_b}{p_b.g \, p_t.g} 
-\frac{m_t^2}{(p_b.g)^2} \Big] \left|D_{--}\right|^2\;,
\eeq
and an equal result from the other gluon polarization, $D^R_{-+-}$. 

This eikonal form is the basis for the counter-term that we use
to subtract the singularities resulting from radiation of a soft
gluon in the top quark decay. We make use of the factorization,
\beq \label{decayfactorization}
\gs^2 C_F \mu^{2 \epsilon} 
\sum_{\lambda_b,\lambda_g,\lambda_t}|{\cD}_{\lambda_b \lambda_g \lambda_t}(p_t,\pw,p_b,p_g) |^2 \to
\sum_{\lambda_b,\lambda_t}|{\cD}_{\lambda_b \lambda_t}|^2(p_t,{\tilde p}_W,{\tilde p}_b) 
\times S(\pt.\pg,\pb.\pg,m_t^2,m_b^2,\pw^2) \; ,
\eeq
where the lowest order matrix element ${\cD}$ is evaluated for momenta
${\tilde p}_W$, ${\tilde p}_b$ of the $W$-boson and $b$-quark, as described in more
detail in Section~\ref{sec:realct}.
In practice, we shall not use the complete eikonal factor as a counterterm, but rather a 
simpler expression that differs from the above by finite terms. Using
\beq
(p_t-p_b-g)^2=\pw^2 \;,
\eeq
and dropping non-singular terms we obtain,
\beq
S(\pt.\pg,\pb.\pg,m_t^2,m_b^2,\pw^2)=
\gs^2 (\mu^2)^\epsilon C_F \times 
\Big[ -\frac{m_b^2}{(p_b.g)^2} +\frac{m_t^2+m_b^2-\pw^2}{p_b.g \, p_t.g} -\frac{m_t^2}{(p_b.g)^2} \Big] \;.
\eeq

The amplitudes for the emission of a gluon from the $\bar t$-$\bar b$ line in anti-top quark decay
are related to the amplitudes presented above, in a similar manner to the lowest order case.
We have,
\beq
\overline{\cal D}^R_{\lambda_a \lambda_g \lambda_c} =
 {\cal C} \left[ {\cal D}^R_{-\lambda_b \, -\lambda_g \, -\lambda_t}\right] \;,
\label{eq:DbarRDRrelation}
\eeq
where the operation ${\cal C}$ is defined in Eq.~(\ref{eq:Cdefn}). We note that, since these
amplitudes contain an additional gluon, this relation is different by an overall sign from
those presented in Eqs.~(\ref{eq:DbarDrelation}) and~(\ref{eq:DbarVDVrelation}).

\subsection{Top decay with virtual gluon radiation from the decay products of the $W$}
\begin{figure}
\begin{center}
\includegraphics[angle=270,width=14cm]{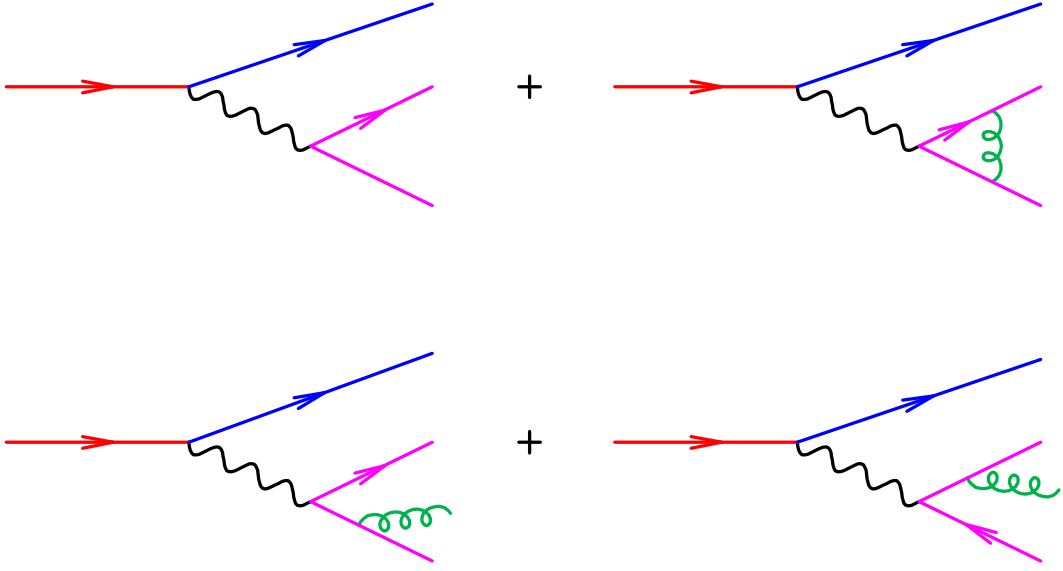}
\caption{Radiative corrections due to the interaction of the decay products of the $W$ in top decay.
The upper line shows the virtual corrections and the lower line the real corrections.}
\label{tdecaywg}
\end{center}
\end{figure}

If the $W$ boson decays hadronically,
\beq
t(p_t) \to b(p_b) + q(q)+\qb(\qb) \;,
\label{eq:tdecayhad}
\eeq
then we should also include the next-to-leading order corrections
to that decay process, as depicted in Figure~\ref{tdecaywg}.
The result for the virtual corrections to the decay of a vector boson 
into a pair of massless fermions is well known~\cite{Altarelli:1979ub}. We can write the form factor as,
\beq
{\cal F}_{\mu} ({\qb},q)=\frac{\gW}{\sqrt{2}} \bar{u}({\qb}) \Gamma_{\mu}({\qb},q) u(q) \;,
\qquad q^2={\qb}^2=0 \;,
\eeq
where the expansion is very simple compared to the general case (c.f. Eq.~(\ref{eq:massiveF})),
\beq
\Gamma^\mu({\qb},q)=\gamma^\mu \gamma_L \Big[ 1+g^2 \; C_F \cg C^L\Big] +O(\ep) \;.
\eeq
Stripping overall colour-factors as usual,
\beq
D^{V_W}_{\lambda_b \lambda_t }= \delta_{\jb \jt}\,
\frac{\gW^2 \gs^2 \cg C_F}{D_W(s_{q\qb})} \Big(\frac{\mu^2}{-s_{q\qb}-i \varepsilon}\Big)^\ep \, {\cD}^{V_W}_{\lambda_b \lambda_t }
\eeq
all of the amplitudes vanish except one,
\beqn \label{virttopdecayWG}
{\cD}^{V_W}_{--} &=& C^L \langle b q \rangle [\qb t] \;.
\eeqn
Note that, for the labelling of momenta indicated in Eq.~(\ref{eq:tdecayhad}), the massless momenta
$t$ and $b$ are defined by choosing $\eta_t = \qb$ and $\eta_b=q$ in Eq.~(\ref{eq:tandbdefn}).
In the four-dimensional helicity scheme the coefficient $C^L$ takes the familiar form,
\beq
C^L=-\frac{2}{\epsilon^2} -\frac{3}{\epsilon} - 7 \;.
\eeq
The amplitudes for the related charge-conjugate process,
\beq
{\bar t}(p_a) \to {\bar b}(p_c) + q(q)+\qb(\qb) \;,
\eeq
can be obtained by a symmetry operation that differs from that in Eq.~(\ref{eq:Cdefn}) only
by a relabelling of momenta. We have,
\beq
\overline{\cal D}^{V_W}_{\lambda_a \lambda_c} = - {\cal C}_W \left[ {\cal D}^{V_W}_{-\lambda_b \, -\lambda_t}\right] \;.
\label{eq:DbarVWDVWrelation}
\eeq
where ${\cal C}_W$ is defined by,
\beq
{\cal C}_W: \qquad t \to a \;, \quad b \to c \;, \quad q \to \qb \;, \quad \qb \to q \;,
 \quad \spa i.j \leftrightarrow \spb i.j \;,
\label{eq:CWdefn}
\eeq
The massless momenta $a$ and $c$ are defined by the choices $\eta_a = q$ and $\eta_c=\qb$ in Eq.~(\ref{eq:aandcdefn}).

\subsection{Top decay with real gluon radiation from the decay products of the $W$}
The diagrams for real radiation of a gluon in the $W$ decay are shown in the bottom row of Figure~\ref{tdecaywg}.
After removing overall factors,
\beq
D^{R_W}_{\lambda_b \lambda_g \lambda_t }= 
\gs \mu^\ep  (t^A)_{i_q i_{\bar{q}}}\, \frac{\gW^2}{D_W(s_{q\qb g})} \,  {\cD}^{R_W}_{\lambda_b \lambda_g \lambda_t} \;,
\eeq
a simple calculation yields the amplitudes,
\beqn
      {\cD}^{R_W}_{---} & = & - \frac{\spb{\qb}.{t} \spab{b}.{\pw}.{\qb}}{\spb{q}.{g}\spb{\qb}.{g}} \;, \nn \\
      {\cD}^{R_W}_{-+-} & = & - \frac{\spa{q}.{b} \spab{q}.{\pw}.{t}}{\spa{q}.{g} \spa{\qb}.{g}}  \;, \nn \\
      {\cD}^{R_W}_{+--} & = &  m_b  \frac{\spa{q}.{g} \spb{\qb}.{t}}{\spa{q}.{b} \spb{q}.{g} } \;, \nn \\
      {\cD}^{R_W}_{++-} & = &0 \;, \nn \\
      {\cD}^{R_W}_{--+} & = &0 \;, \nn \\
      {\cD}^{R_W}_{-++} & = &  -m_t  \frac{\spa{q}.{b} \spb{\qb}.{g}}{\spa{\qb}.{g} \spb{\qb}.{t}} \;, \nn \\
      {\cD}^{R_W}_{+-+} & = &0 \;, \nn \\
      {\cD}^{R_W}_{+++} & = &0 \;,
\eeqn
where $p_W = p_q + p_{\qb} + p_g$.
The counterterm in this case is the standard Catani-Seymour dipole associated with 
final-final radiation from massless partons~\cite{Catani:1996vz}.

The equivalent amplitudes for the anti-top quark decay process are given by,
\beq
\overline{\cal D}^{R_W}_{\lambda_a \lambda_g \lambda_c} =
 {\cal C}_W \left[ {\cal D}^{R_W}_{-\lambda_b \, -\lambda_g \, -\lambda_t}\right] \;,
\label{eq:DbarRWDRWrelation}
\eeq
where the operation ${\cal C}_W$ is defined in Eq.~(\ref{eq:CWdefn}). Once again this relation
has the opposite sign from that in Eq.~(\ref{eq:DbarVWDVWrelation}).

\section{Counter-term for real radiation}
\label{sec:realct}
In this section we describe the implementation of the counter-term
that is used to cancel the soft singularity resulting from real
radiation from the $t$-$b$ line in the decay of the top quark. The counter-term has
already been introduced in Eq.~(\ref{decayfactorization}), which
makes clear that the lowest order matrix element is evaluated at
momenta ${\tilde p}_W$ and ${\tilde p}_b$ of the $W$ boson and bottom quark respectively.
These momenta are related to those for which the real radiation matrix element is evaluated
($p_W$, $p_b$, $p_g$) as follows. 

The transformed momenta ${\tilde p}_W$ and ${\tilde p}_b$ are generated by a Lorentz
boost along the direction of the $W$ in the top rest frame, following the method
of ref.~\cite{Campbell:2004ch}~(Section IV). The transformation is generalized slightly
in order to incorporate the effect of a non-zero bottom quark mass. Thus we still have
${\tilde p}_W^\mu = \Lambda^\mu_{~\nu} \, \pw^\nu$ with,
\beqn
\Lambda^{\mu\nu}&=&g_{\mu \nu}
     +\frac{\sinh(x)}{\sqrt{(\pt \cdot \pw)^2 -\pw^2 \pt^2}} \; 
\Big(\pt^{\mu} \pw^{\nu}-\pw^{\mu} \pt^{\nu} \Big) \nn  \\
 &+&\frac{\cosh(x)-1}{(\pt \cdot \pw)^2 -\pw^2 \pt^2} \;
      \Big(\pt \cdot \pw \; (\pt^\mu \pw^\nu+\pw^{\mu} \pt^{\nu})
    -\pw^2 \; \pt^{\mu} \pt^{\nu}-\pt^2 \; \pw^{\mu} \pw^{\nu}\Big)
\label{eq:lorentztrans}
\eeqn
but ${\tilde p}_b$ is now constrained by ${\tilde p}_b^2 = (p_t-{\tilde p}_W)^2 = m_b^2$.
This constraint determines the coefficients that appear in the transformation,
\beqn
\label{eq:lorentztransa}
\sinh(x)&=&\frac{1}{2 \; \pt^2 \pw^2} \Big[-\sqrt{\lambda(\pt^2,\pw^2,\pb^2)} \, \pt\cdot \pw
 +(\pt^2+\pw^2-\pb^2) \sqrt{(\pt \cdot \pw)^2 -\pw^2 \pt^2}\; \Big] \;,
 \nonumber \\
\cosh(x)&=&\frac{1}{2 \; \pt^2 \pw^2} \Big[+(\pt^2+\pw^2-\pb^2) \pt\cdot \pw
 -\sqrt{\lambda(\pt^2,\pw^2,\pb^2)} \, \sqrt{(\pt \cdot \pw)^2 -\pw^2 \pt^2}\; \Big] \;.
\eeqn
This Lorentz transformation is also used to determine the modified momenta of the $W$ decay products.

With the subtraction fully specified, we now turn to the issue of performing the
integration of this counter-term analytically in order that the poles can be extracted
and cancelled against those appearing in the virtual calculation.
From Eq.~(\ref{eq:PSfactorized}) we may write the phase space
for the decay of an on-shell top quark as 
\beqn \label{eq:phasespacefact}
d\Phi^{(3)} (p_t; \pw, \pb, \pg)
&=& d \Phi^{(2)} (\pt; \pw, \pb) \times [dg(\pt ,\pw, z)]  \nn \\
&\equiv& d \Phi^{(2)} (\pt; \tpW, \tpb) \times [dg(\pt ,\tpW, z)] \;.
\eeqn
The equivalence in Eq.~(\ref{eq:phasespacefact}) 
follows from Eq.~(\ref{eq:PS2}) in the appendix because
$d ^{n-1}\om_w= d ^{n-1}\om_{\tilde w}$ since $\pw$ and ${\tpW}$
are related by a boost. 

The counter-term may now be integrated in exactly the same manner
as the real radiation matrix elements that enter the calculation of the
total width, as detailed in Appendix~\ref{sec:totalwidth}.
The integrated counter-term corresponds to the sum of the integrals
$S_1$, $S_2$ and $S_3$, expressions for which are given in
Eqs.~(\ref{eq:S1res}),~(\ref{eq:S2res}) and (\ref{eq:S3res}).
We thus arrive at the final expression for the integrated counter-term,
\beqn
&&\int [dg(\pt,{\tpW},z)]\, S(\pt.\pg,\pb.\pg,m_t^2,m_b^2,\pw^2) =  S_1+S_2+S_3 = \nonumber \\
&&
2 g^2 \cg C_F\; \left(\frac{\mu^2}{m_t^2}\right)^\ep
       \Bigg\{ \left[\frac{2}{\ep} - 4\ln\left(\frac{4 \P3b^2}{\om \beta}\right)\right]
        \Big(1-\frac{\Pzb}{\P3b} \Ypb\Big) +4
          + \frac{2}{\P3b}\Big((1-\om^2) \Ypb
          +(1-\beta^2)\Ywb \Big) \nn \\
         &+&\frac{\Pzb}{\P3b} 
     \Big[2\Ypb - 6\Ypb^2 + 4\Ywb\ln(\beta) - 6\, \li\left(1-\frac{\Pmb}{\Ppb}\right)
     -2 \, \li\left(1-\Ppb\right)+2 \, \li\left(1-\Pmb\right)\Big] \Bigg\} \;.
\eeqn

\section{Consistent treatment of top quark decay in perturbation theory}
\label{Radiation_in_decay}
Since we are everywhere treating the top quark as on shell, the full cross
section integrated over the decay products of the top 
will be given by the production cross section 
multiplied by the branching fraction to the chosen decay channel. 
This statement should hold at all orders in perturbation theory.
For example, we can write the differential NLO cross section
for single top production followed by semi-leptonic decay schematically as,
\beq \label{prodanddecay}
\sigma^{NLO}(p p \to t(\to \nu e^+  b) +X)  = (\sigma_0+\as \sigma_1) \times 
 \frac{d\Gamma^{(l)}_0+\as d \Gamma^{(l)}_1}{\Gamma_0+\as  \Gamma_1} \;,
\eeq
where $\sigma_0$, $\Gamma^{(l)}_0$, $\Gamma_0$ are the lowest order contributions 
to the production rate, semileptonic decay width $\Gamma(t \to \nu e^+  b)$, and total top width
and $\as \sigma_1$, $\as \Gamma^{(l)}_1$ and
$\as \Gamma_1$ the corresponding NLO corrections.
The total top width is given by the sum of the partial widths to the various decay channels,
\beq
\Gamma_t = \sum_{i,j} \Gamma( t \to f_i {\bar f}_j b + X) \;,
\eeq
which factorizes in the narrow width approximation for the $W$-boson to,
\beq
\Gamma_t \to \Gamma( t \to b W + X)
 \times \sum_{i,j} \frac{ \Gamma( W \to f_i {\bar f}_j)}{\Gamma_W} \equiv \Gamma( t \to b W + X) \;.
\eeq
The quantity $\Gamma( t \to b W + X)$ has the perturbative expansion,
 $\Gamma( t \to b W + X) = \Gamma_0 + \as \Gamma_1$, where 
$\Gamma_0,\Gamma_1$ are as given in Eqs.~(\ref{eq:LOwidth}),~(\ref{totalwidth}).
This is the expression for the total width used in the denominator of Eq.~(\ref{prodanddecay}).

In order to include the effect of radiation in the decay of the top quark properly,
we must be careful to ensure that we perform the perturbative expansion
in the strong coupling in a consistent manner. 
As written in Eq.~(\ref{prodanddecay}) the NLO calculation includes 
a contribution of relative order $\as^2$ corresponding
to corrections in both production and decay stages simultaneously. 
We follow the treatment in ref.~\cite{Melnikov:2009dn} and simply expand
Eq.(\ref{prodanddecay}) in $\as$, discarding terms of order $\as^2$ or higher,
\beqn
\sigma^{NLO}
&=& \sigma_0\times \frac{d\Gamma^{(l)}_0}{\Gamma_0}
+ \sigma_0\times \frac{\as d \Gamma^{(l)}_1}{\Gamma_0}
 +\as \sigma_1 \times \frac{d \Gamma^{(l)}_0}{\Gamma_0}
 -\as \sigma_0 \times \frac{d \Gamma^{(l)}_0}{\Gamma_0}\frac{\Gamma_1}{\Gamma_0} \;.
\label{eq:widthcorr}
\eeqn
Performing the full integration over the final state, i.e. making the substitutions
$d\Gamma^{(l)}_0 \to \Gamma^{(l)}_0$ and $d\Gamma^{(l)}_1 \to \Gamma^{(l)}_1$, 
we recover the NLO cross section for corrections in production only, 
\beqn \label{eq:sigmatimesBRunchanged}
\sigma^{NLO} &=& (\sigma_0 + \as \sigma_1)\times Br(W \to \nu e^+) +\as \sigma_0 Br(W \to \nu e^+) \left[
\frac{\Gamma^{(l)}_1}{\Gamma^{(l)}_0}-\frac{\Gamma_1}{\Gamma_0}\right] \nonumber \\
&\equiv & (\sigma_0 + \as \sigma_1)\times Br(W \to \nu e^+)
\eeqn
where the equality follows, for example in the narrow width approximation, because,
\beq
\frac{\Gamma^{(l)}_1}{\Gamma^{(l)}_0} = 
 \frac{\Gamma_1(t \to bW+X) Br(W \to \nu e^+)}{\Gamma_0(t \to bW+X) Br(W \to \nu e^+)} \;.
\eeq
We note that while Eq.~(\ref{eq:sigmatimesBRunchanged}) 
is a desirable outcome, it is only true in the case that all
degrees of freedom associated with the decay are completely integrated out. In the
presence of experimental cuts, for instance on the leptons or $b$-quarks present in the
decay, the two calculations (i.e.\ with and without radiation in decay), 
can predict different cross sections.
The above discussion pertains to the case of NLO corrections to the production
and leptonic decay of a single top quark. For top pair production, with both top
quarks decaying leptonically, the last term that appears in
Eq.~(\ref{eq:widthcorr}) appears once for each quark so that this correction factor is
doubled.

We note that since the top quark is produced exactly on-shell, the value of the top quark
width is important only to ensure that we obtain the correct branching ratio when
including the decay. This can clearly be achieved by using 
$\Gamma_0$ as shown in Eq.~(\ref{eq:widthcorr}). This is therefore the value of the width
used in our code.

The argument given above also holds, {\it mutatis mutandis}, for hadronic
instead of leptonic decays of the $W$-boson.

\section{Phenomenology}
\label{pheno}

\subsection{Input parameters}

The results presented in this paper are obtained with the latest version of
the MCFM code (v6.2).
The electroweak parameters that we regard as inputs are,
\begin{eqnarray}
M_W &=& 80.398~\mbox{GeV}\;, \;\; \Gamma_W = 2.1054~\mbox{GeV} \;, \\
G_F &=& 1.16639 \times 10^{-5} \, \mbox{GeV}^{-2}\;.
\end{eqnarray}
The top and bottom quarks have the pole masses,
\beq
m_t = 172.5~\mbox{GeV}\;, \;\; m_b = 4.7~\mbox{GeV} \;.
\eeq

For the parton distribution functions (pdfs) we use the sets of Martin, Stirling,
Thorne and Watt~\cite{Martin:2009iq}. For the calculation of the LO results presented here we employ
the corresponding LO pdf fit, with 1-loop running of the strong coupling and
$\alpha_s(M_Z)=0.13939$. Similarly, at NLO we use the NLO pdf fit, with
$\alpha_s(M_Z)=0.12018$ and 2-loop running.

In our calculations of the top pair and $s$-channel single top processes
we set the factorization and renormalization scales equal to the top
quark mass. For the $t$-channel process we employ two scales, one for
evaluating contributions associated with the light quark line ($\mu_l$)
and the other for the heavy quark line ($\mu_h$).
We set $\mu_l = m_t/2$ and $\mu_h=m_t/4$, as advocated in
Ref.~\cite{Campbell:2009gj}.

%%%%%%%%%%%% S-CHANNEL SINGLE TOP %%%%%%%%%%%%%%%%%%%%%%%%%%%%%%%%%%%%%%%%%%%%%%%%

\subsection{$s$-channel single top at the Tevatron}
In order to illustrate some of the features of our calculation, we begin by
considering the $s$-channel single top process. Although the thrust of this paper
is the inclusion of the top quark decay allowing realistic experimental cuts
to be applied to the top decay products, it it instructive to first consider
the case in which no cuts are applied.

The predicted cross section for the process $p\bar p \to t(\to \nu e^+ b) \bar b$
at the Tevatron, computed at various levels of sophistication, is shown in
Table~\ref{table:schanxsecnocuts}. In this table we present results of performing
the calculation at LO, at NLO including radiation in production only, and at NLO
including radiation in both production and decay. Moreover, we consider three
different kinematic approximations: setting $m_b=0$ and using the narrow-width
approximation for the $W$-boson (equivalent to the approach in Ref.~\cite{Campbell:2004ch}),
using $m_b=4.7$~GeV but still using the narrow-width approach, and finally
using $m_b=4.7$~GeV and implementing the full Breit-Wigner form of the $W$ propagator.
\begin{table}
\begin{center}
\begin{tabular}{|l|c|c|c|}
\hline
Treatment of $W$-boson and $b$-quark &~~$\sigma_{LO}$~[fb]~~ & $\sigma_{NLO}$ (prod.)~[fb] &  $\sigma_{NLO}$ (prod.+decay)~[fb] \\
\hline
Narrow width, $m_b=0$       & 30.98(2) & 48.84(3) & 48.82(3) \\
Narrow width, $m_b=4.7$~GeV & 30.78(2) & 48.61(3) & 48.60(3)\\
Breit-Wigner, $m_b=4.7$~GeV & 30.77(2) & 48.61(3) & 48.59(3)\\
\hline
\end{tabular}
\caption{Cross sections in femtobarns for the $s$-channel single top process, computed
with various levels of sophistication. No cuts have been applied to the final state.
The numerical integration errors are shown in brackets.} 
\label{table:schanxsecnocuts}
\end{center}
\end{table}
We first observe that, by construction, the two columns of calculated NLO cross sections
are in perfect agreement, thanks to the presence of the last term in Eq.~(\ref{eq:widthcorr}).
Moreover, the slight adjustment to the width of the $W$-boson when using the Breit-Wigner rather
than the narrow-width approximation, as indicated in Table~\ref{widthvalues}, ensures that the
final two rows of Table~\ref{table:schanxsecnocuts} are also in excellent agreement.
Finally, we observe that the effect of including the $b$-mass in this process is very small,
resulting in a decreases in the cross-section by approximately 0.5\%. We note that this effect
is not the result of including the mass of the $b$-quark in the top quark decay but is instead
due to the treatment of the $\bar b$ quark.

We now perform the same analysis using a realistic set of experimental cuts. In particular we
use the cuts employed in a recent search for the Higgs boson using the $WH(\to b\bar b$ associated
production channel, for which $s$-channel single top production is an irreducible background~\cite{CDFNote10796}.
All jets (both light and heavy flavour) are
clustered according to the anti-$k_T$ algorithm with a distance parameter
$D=0.7$. We require that the algorithm finds at least two jets that satisfy,
\beq
p_T(\mbox{jet}) > 20~\mbox{GeV} \;, \qquad y(\mbox{jet}) < 2 \;.
\eeq
and the charged lepton acceptance is defined by,
\beq
p_T(\mbox{lepton}) > 20~\mbox{GeV} \;, \qquad y(\mbox{lepton}) < 1 \;.
\eeq
In addition we require that there is at least $20$~GeV of missing transverse momentum.
Our results for the process $p\bar p \to t(\to \nu e^+ b) \bar b$ under this set of cuts
are shown in Table~\ref{table:schanxsec}.
\begin{table}
\begin{center}
\begin{tabular}{|l|c|c|c|}
\hline
Treatment of $W$-boson and $b$-quark & ~~$\sigma_{LO}$~[fb]~~ & $\sigma_{NLO}$ (prod.)~[fb] &  $\sigma_{NLO}$ (prod.+decay)~[fb] \\
\hline
Narrow width, $m_b=0$       & 12.14(2) & 19.96(2) & 20.03(2) \\
Narrow width, $m_b=4.7$~GeV & 12.12(2) & 19.96(2) & 20.01(2) \\
Breit-Wigner, $m_b=4.7$~GeV & 12.08(2) & 19.88(2) & 19.95(2) \\
\hline
\end{tabular}
\caption{Cross sections in femtobarns for the $s$-channel single top process, computed
with various levels of sophistication. The cuts appropriate for the Higgs search,
as described in the text, have been applied.} 
\label{table:schanxsec}
\end{center}
\end{table}
As anticipated, the application of cuts results in a small difference between
the two columns of NLO results. We find that the NLO cross section including radiation
in production and decay stages is approximately $0.3\%$ higher than the result for
radiation in production only. For similar reasons, the cuts also induce a slight
reduction in the cross section (by less than $0.5\%$) when working with the
full Breit-Wigner propagator rather than in the narrow width approximation.
We also see that the effect of including the $b$-mass is negligible, due to the
requirement that at least two jets are reconstructed with transverse momenta that
are relatively large compared to the $b$-quark mass.

%%%%%%%%%%%% T-CHANNEL SINGLE TOP %%%%%%%%%%%%%%%%%%%%%%%%%%%%%%%%%%%%%%%%%%%%%%%%

\subsection{$t$-channel single top at the LHC}
The production of single top quarks via the $t$-channel process is the dominant
production mode at the LHC. The final state is hard to distinguish from top pair
production and a variety of QCD backgrounds such as the production of a $W$ boson
in association with jets. In order to distinguish
single top events from the main background processes, a recent CMS
analysis~\cite{Chatrchyan:2011vp} relies on accurate theoretical predictions for
two observables.

The first is the rapidity of the
light jet that is present in the event. Since at LO this light jet is produced
by the $t$-channel emission of a $W$ boson from an initial quark, it is found
primarily in the forward direction, in contrast to the centrally-produced
light jets from background processes.
The second observable,  $\cos\theta^\star$ is computed by performing a boost to the rest
frame of the reconstructed top quark and then defining $\theta^\star$ as the angle
between the charged lepton and the light jet in that frame.
For both these observables it is interesting to investigate
to what extent the distributions are modified by higher order
corrections to both the production and decay processes.

For the study presented here we have adopted the cuts from Ref.~\cite{Chatrchyan:2011vp}.
Light and heavy-flavour jets are
clustered according to the anti-$k_T$ algorithm with a distance parameter
$D=0.5$ and only satisfy a fairly loose set of cuts,
\beq
p_T(\mbox{jet}) > 30~\mbox{GeV} \;, \qquad y(\mbox{jet}) < 5 \;.
\eeq
The cuts on the charged lepton are,
\beq
p_T(\mbox{lepton}) > 20~\mbox{GeV} \;, \qquad y(\mbox{lepton}) < 2.1 \;,
\eeq
and no requirement is made on the missing transverse momentum.
Following the CMS analysis, we demand that exactly two jets are present
after jet clustering: the appropriately-charged bottom jet and a light
(non-tagged) jet. We reconstruct the top quark by combining the leptonically
decaying $W$ boson (reconstructed perfectly, by assumption) with the $b$-jet.
If the invariant mass of the $W$-$b$-light-jet system is closer to $m_t$ than
the invariant mass of the $W$-$b$ system then the radiation is assumed to occur
in the decay and the event is dropped.

\begin{figure}[ht]
\begin{center}
\includegraphics[angle=0,width=7.5cm]{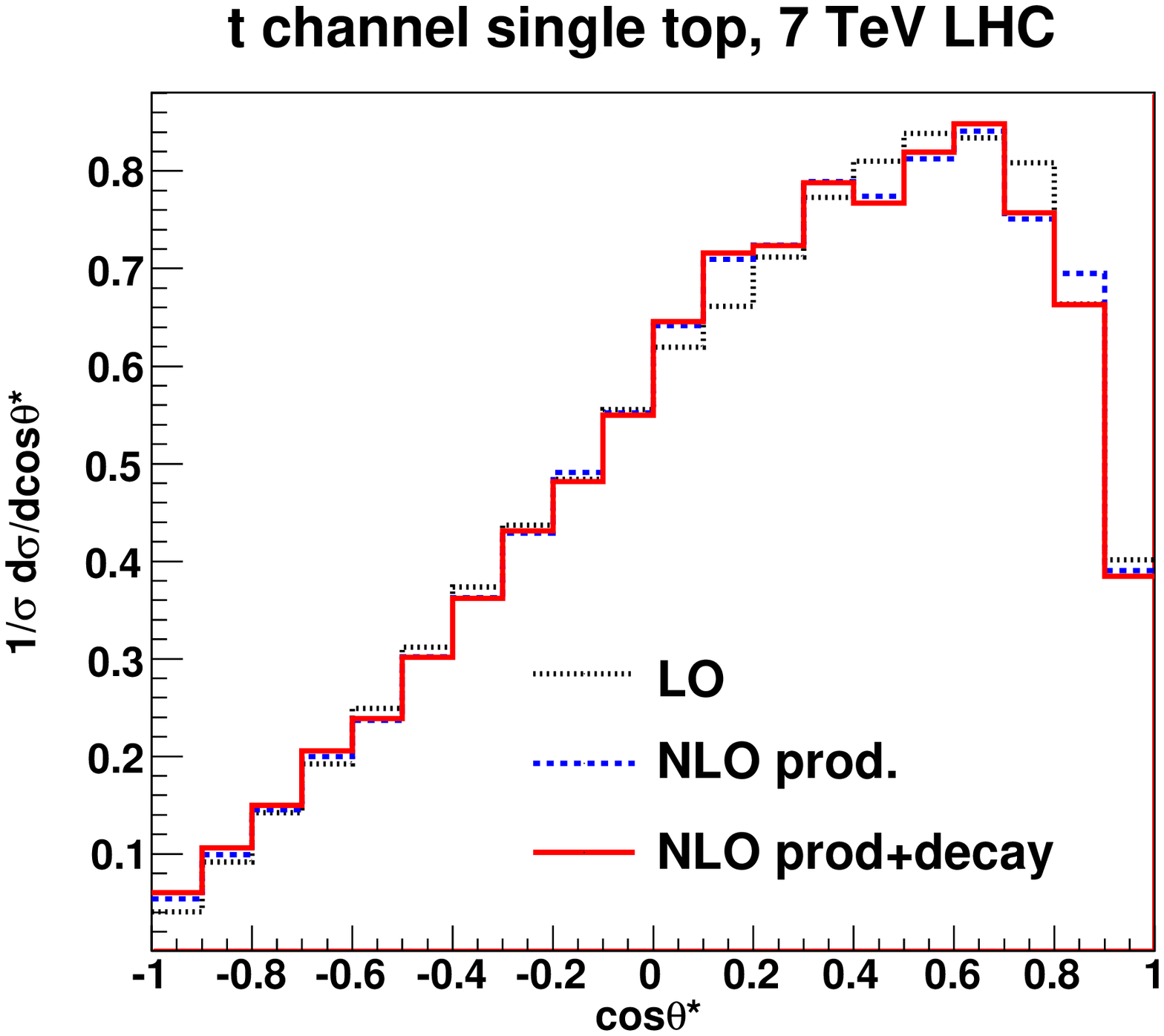}
\includegraphics[angle=0,width=7.5cm]{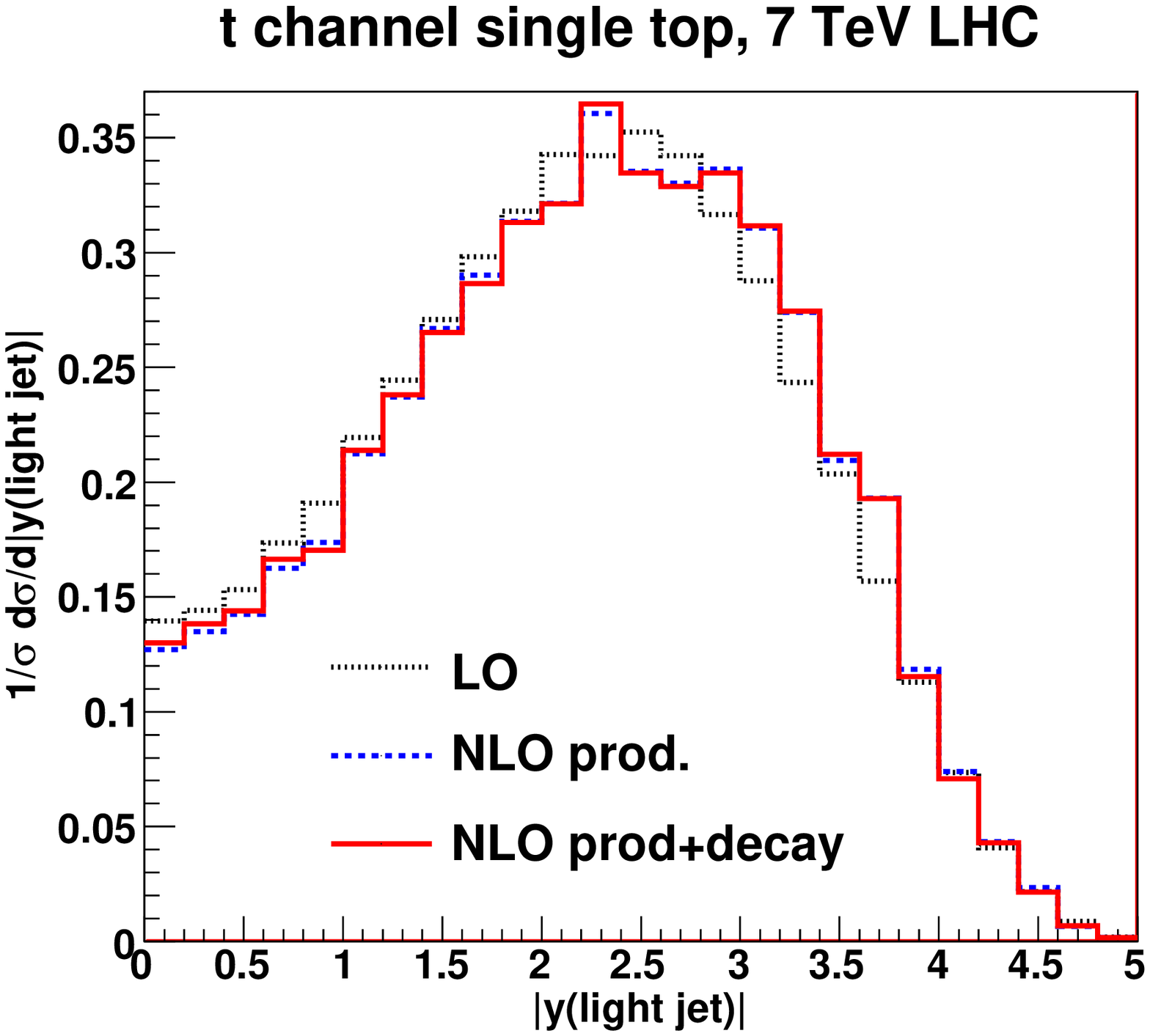}
\caption{Distributions of $\cos\theta^*$ (defined in the text) and $|y_{\rm light}$ at the
7 TeV LHC, computed using the set of cuts described in the text and adapted from
Ref.~\cite{Chatrchyan:2011vp}.}
\label{fig:tchanplots}
\end{center}
\end{figure}

Our results are shown in Figure~\ref{fig:tchanplots}.
We observe that the effect of the NLO corrections in the production is a slight change
in the shape of both distributions. However, the effect of additionally including the
radiation in the decay is not significant. 

%%%%%%%%%%%% TOP PAIRS %%%%%%%%%%%%%%%%%%%%%%%%%%%%%%%%%%%%%%%%%%%%%%%%

\subsection{Top pair phenomenology}

The phenomenology of top pair production is very rich in comparison with the single top 
processes due to the much higher production cross section. We therefore perform
a more detailed study of top pair production in this section.

\subsubsection{Top production at the LHC}

We first provide predictions for some basic kinematic distributions that have
recently been analyzed by the CMS collaboration~\cite{Chatrchyan:2011yy}. The cuts
employed in our study are the ones presented in that analysis.
Both light and heavy-flavour partons are clustered according to the anti-$k_T$ algorithm
with a distance parameter $D=0.5$ and the resulting jets must satisfy the acceptance cuts,
\beq
p_T(\mbox{jet}) > 30~\mbox{GeV} \;, \qquad y(\mbox{jet}) < 2.4 \;.
\eeq
We require that two $b$-jets are found by the algorithm, out of a total of at least
two jets (for the dilepton case, with the top and anti-top decay leptonically) and at least
four jets (in the lepton+jets case, with one top decaying hadronically).
The cuts on the charged lepton depend on the decay channel:
\beqn
\mbox{dilepton}:    && p_T(\mbox{lepton}) > 20~\mbox{GeV} \;, \qquad y(\mbox{lepton}) < 2.4 \;, \nn \\
\mbox{lepton+jets}: && p_T(\mbox{lepton}) > 30~\mbox{GeV} \;, \qquad y(\mbox{lepton}) < 2.1 \;,
\eeqn
but no requirement is made on the missing transverse momentum.

We will also present distributions for the $W$ bosons, top and anti-top quarks that are reconstructed
according to the following algorithm. For simplicity we assume that a leptonically decaying
$W$ is perfectly reconstructed. For a hadronic $W$ decay we consider as candidates all
pairs of light jets and, if there are three light jets, also the system of all three together (to account for
radiation in the decay). The system whose invariant mass is closest to $m_W$ is assigned
as the hadronic $W$ decay. For the top and anti-top quarks, we consider both the system
consisting of the $W$ and the appropriate bottom quark, and the system that also contains
the remaining light jet (if it was not already assigned to the $W$ decay). The combination
of assignments that results in invariant masses closest to $m_t$ for both top and anti-top quarks
is considered the proper solution. We note that, since we produce both top and anti-top quarks
on-shell, this reconstruction is often perfect. However, in the case that initial state radiation
in the production stage is merged with a bottom quark in the final state, one of the top
quark masses may be reconstructed far from its mass shell. We have not tried to remove such
configurations in the results presented here, although the code is flexible enough to pursue
such approaches.

\begin{figure}[ht]
\begin{center}
\includegraphics[angle=0,width=7.5cm]{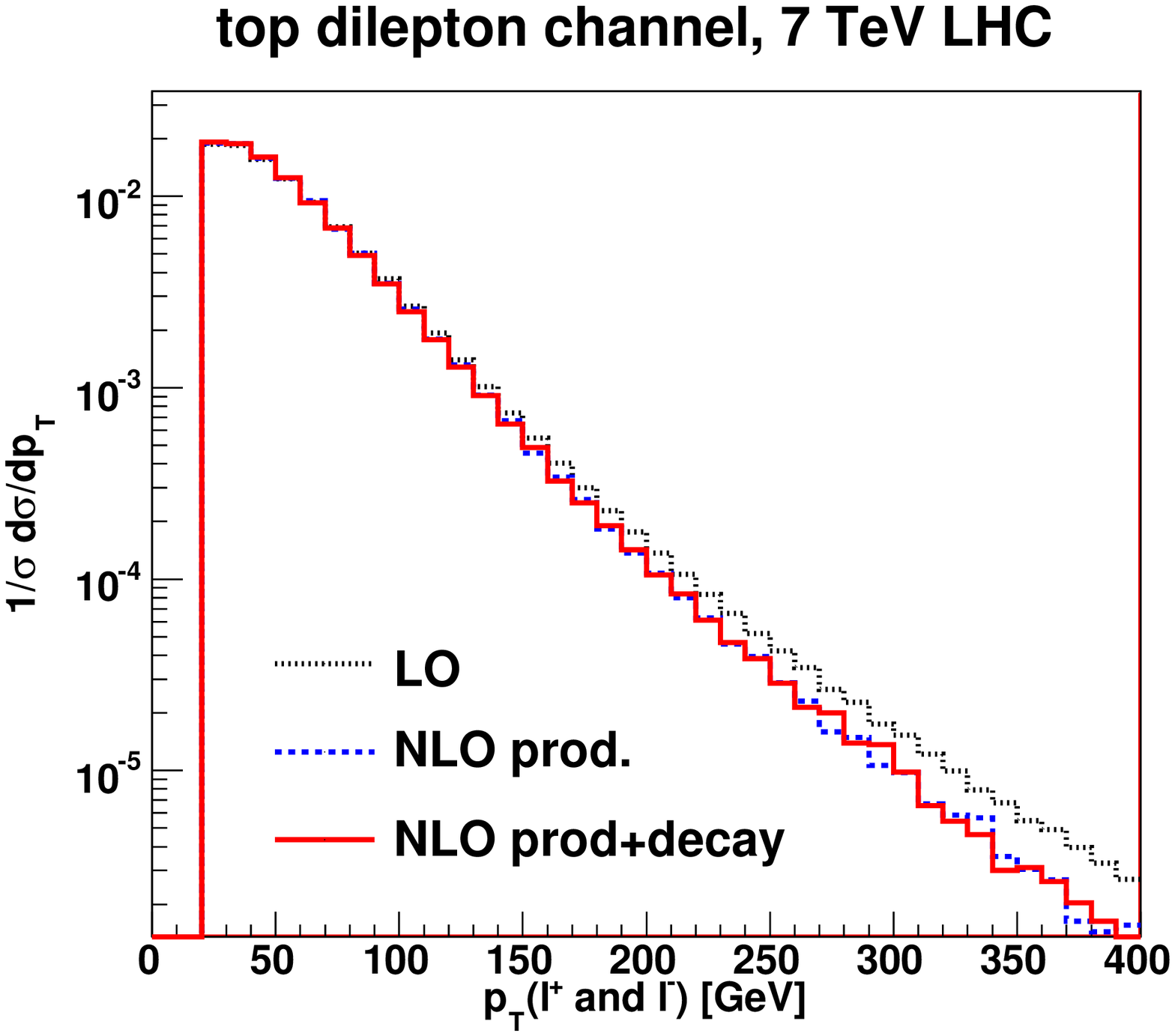}
\includegraphics[angle=0,width=7.5cm]{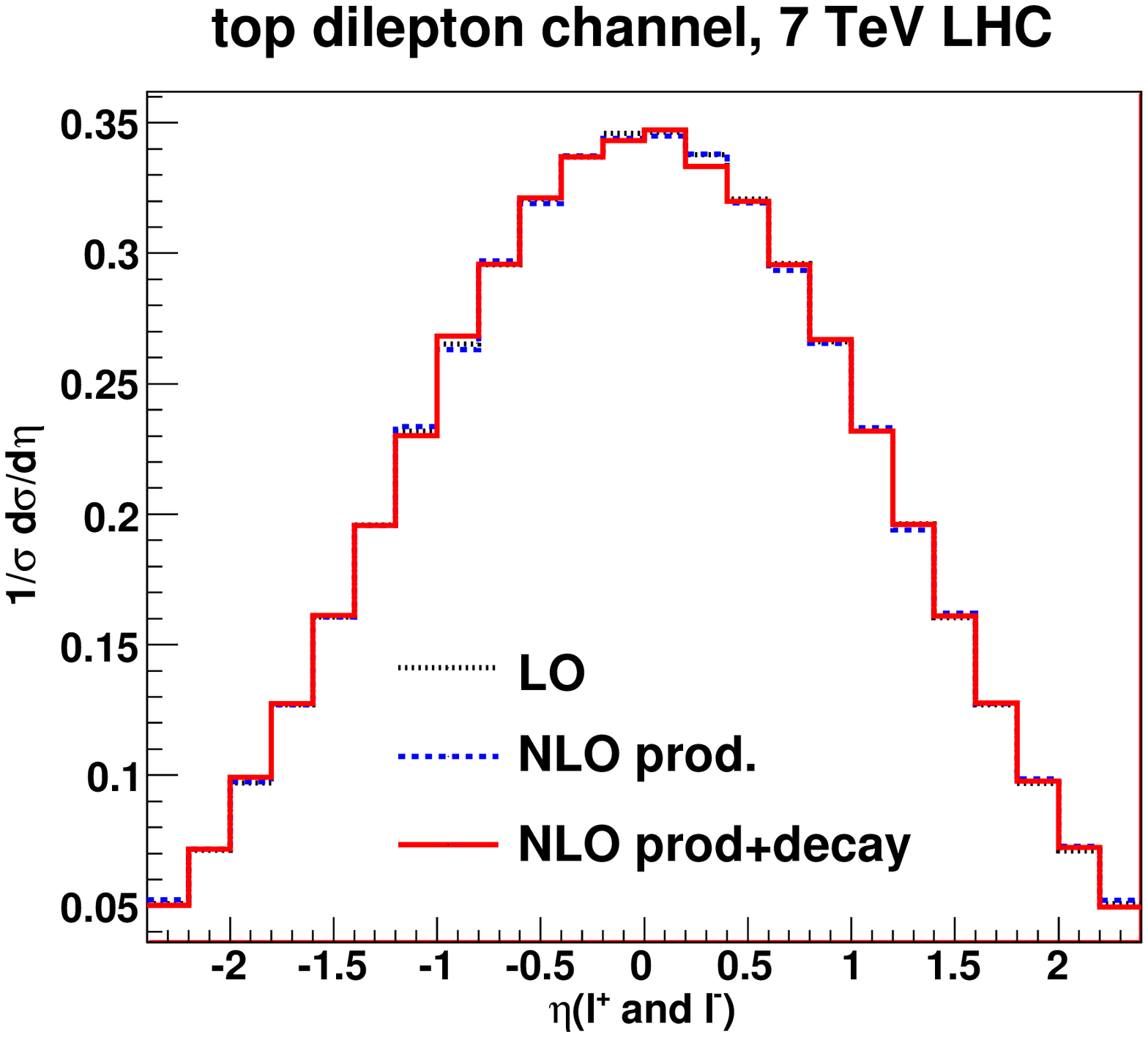} \\ \vspace*{0.5cm}
\includegraphics[angle=0,width=7.5cm]{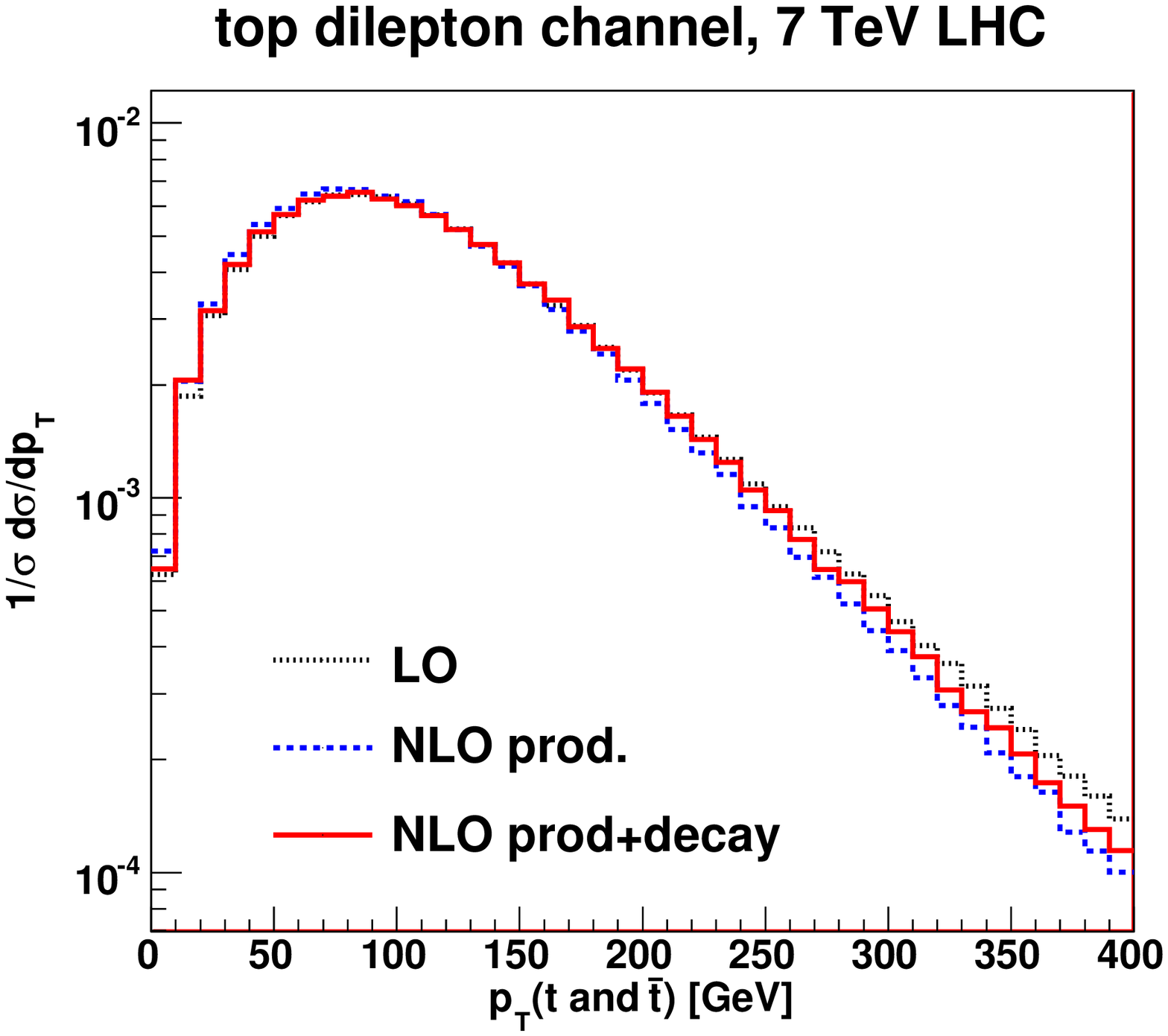}
\includegraphics[angle=0,width=7.5cm]{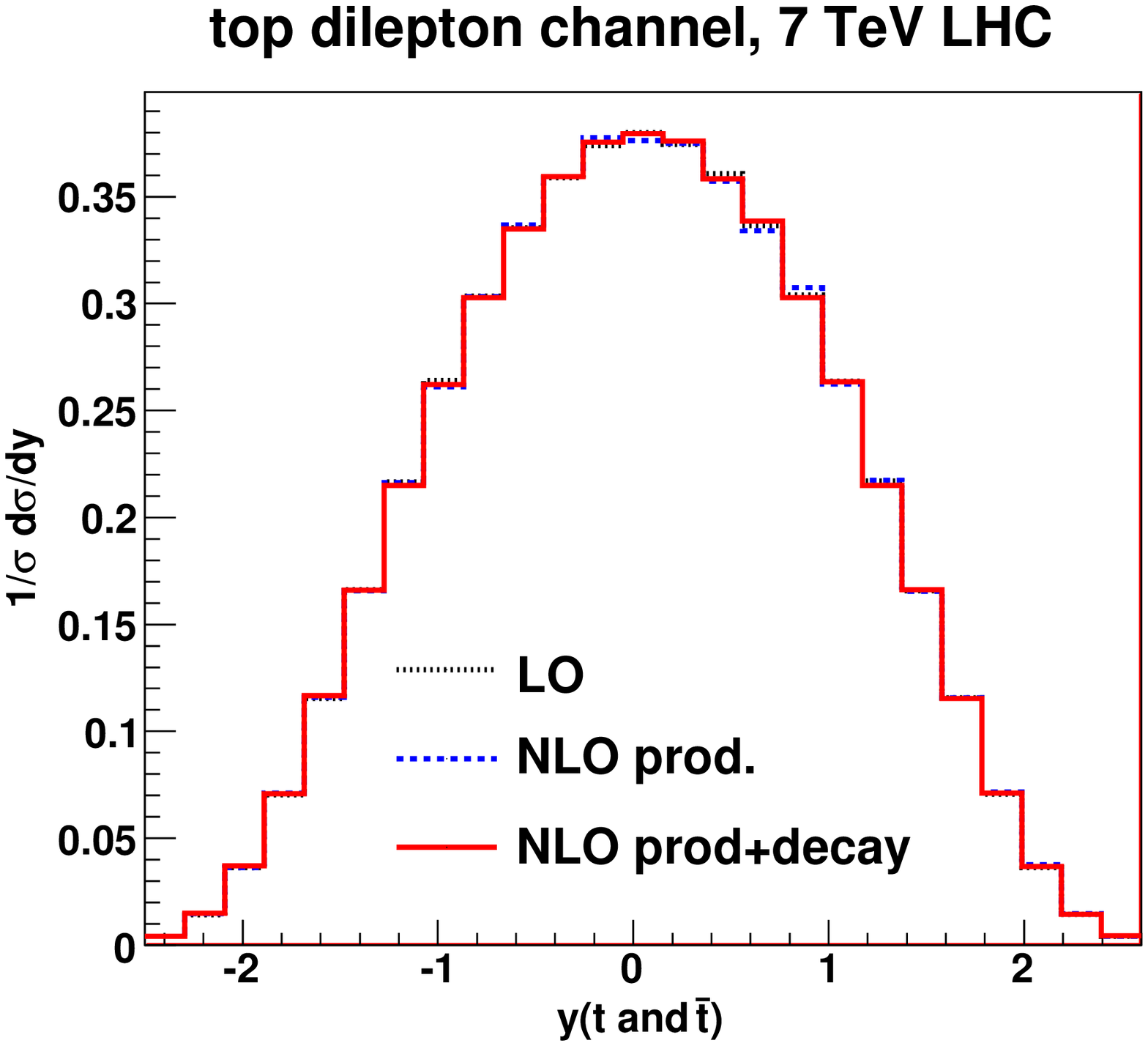}
\caption{Predictions for various observables in the top pair dilepton channel
at 7 TeV LHC, calculated at LO (black), NLO in production (blue) and NLO in production
and decay (red). The observables are: the $p_T$ (top left) and rapidity (top right)
of the charged leptons and the $p_t$ and rapidity of the reconstructed top and anti-top
quarks.  In each case, a single event enters the histograms twice, once each for the
particle ($\ell$ or $t$) and anti-particle ($\bar\ell$ or $\bar t$).}
\label{fig:dileptongen}
\end{center}
\end{figure}

In Figure~\ref{fig:dileptongen} we present a selection of observables for the 
dilepton process. Focusing first on the leptonic observables we see that the NLO
corrections have a considerable impact on the shape of the $p_T$ distribution of the
charged leptons but that their rapidity distribution is left unchanged. Moreover, the
effect of including NLO corrections in the decay of the top quark is negligible.
Turning to the observables obtained after reconstruction of the top and anti-top
quark momenta, we see that the rapidity distribution of the top quarks is not
affected by the QCD corrections. The shape of the transverse momentum distribution
of a single top quark does receive substantial corrections at large $p_T \gsim 250$~GeV.
In addition, the two NLO curves also differ: the prediction when including QCD radiation
in the top decay lies between the LO and NLO (production only) curves. 

\begin{figure}[ht]
\begin{center}
\includegraphics[angle=0,width=7.5cm]{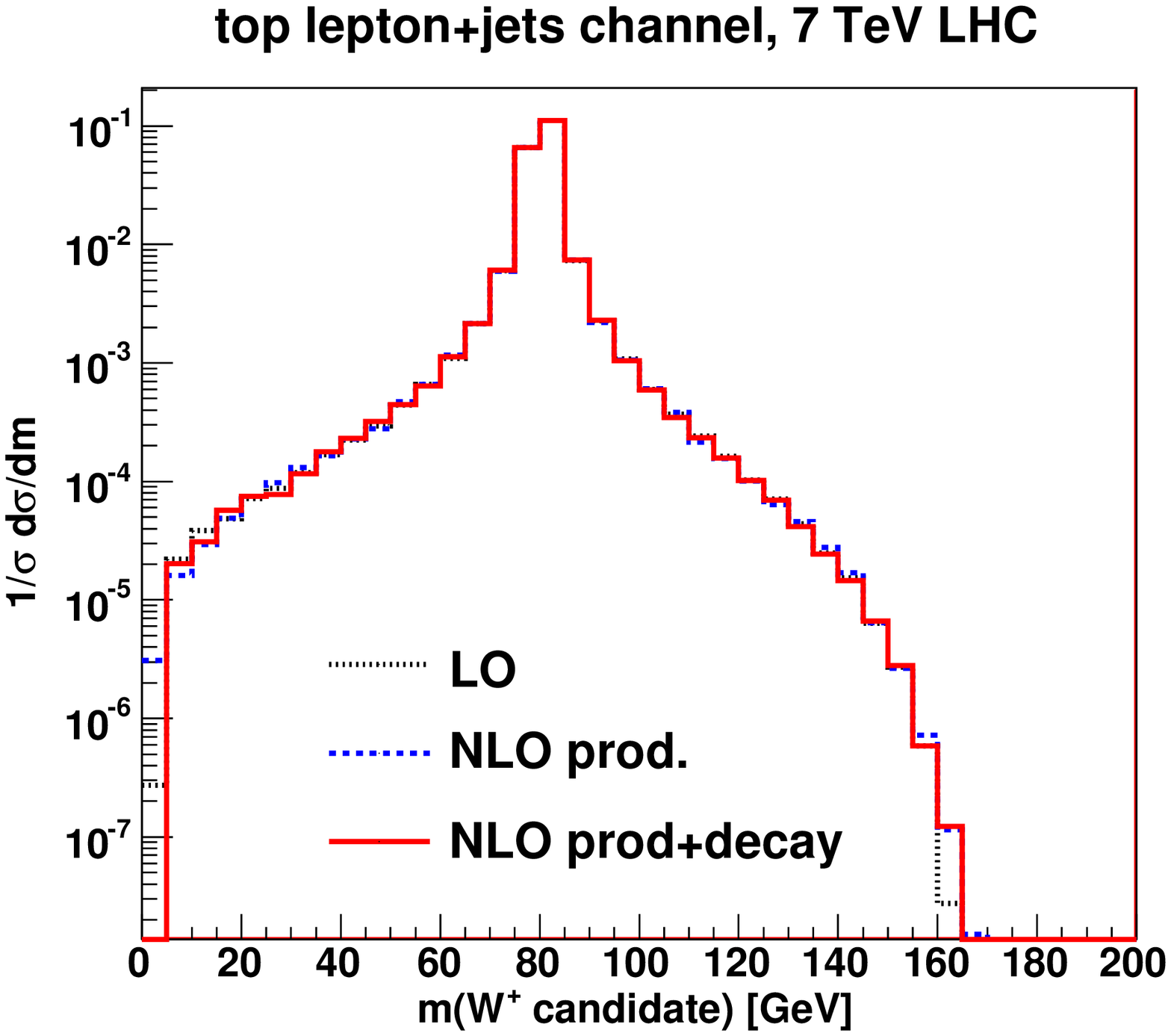}
\includegraphics[angle=0,width=7.5cm]{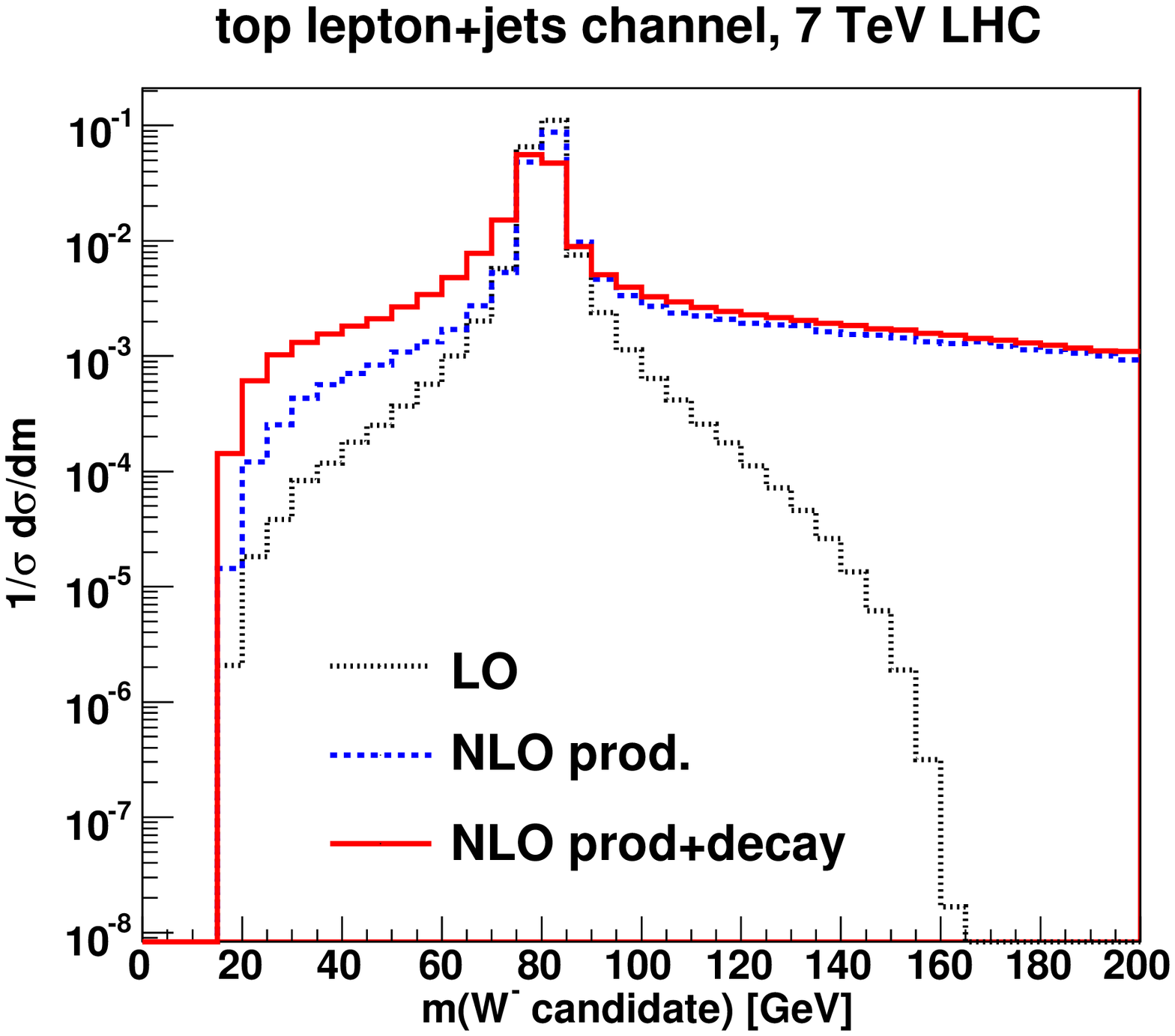}
\caption{Distribution of the $W$-boson candidate invariant masses in the lepton+jets channel
at the 7 TeV LHC. The top quark decays leptonically, so the $W^+$ candidate (left)
is perfectly reconstructed, while the anti-top quark decays hadronically leading to ambiguity
in the reconstruction of the $W^-$-boson at NLO (right).}
\label{fig:leptonjetsWmasses}
\end{center}
\end{figure}

For the lepton+jets channel we consider the process
$pp \to t(\to e^+ \nu b) \bar t(\to e^- \bar\nu \bar b)+X$, i.e. we consider the leptonic
decay of the top quark and the hadronic decay of the anti-top quark.
In Figure~\ref{fig:leptonjetsWmasses} we show the invariant mass spectrum of the
$W^+$ and $W^-$ bosons produced
in this process. We note that for the $W^+$ boson we have made the simplification that
the neutrino 4-momentum is known, while the $W^-$ boson is reconstructed according
to the algorithm above.
We see that, at LO, the predicted distribution for $W^-$ is identical to $W^+$, since
there is no ambiguity in identifying the jets (we assume that the $b$-jets are
perfectly tagged). However, at NLO the additional radiation -- either before or after
the top quark decay -- may be mistakenly assigned to the $W^-$ boson, leading to the
substantial change in shape of the distribution. In addition there is a significant
difference between the two NLO predictions, with radiation in the decay leading to
a bigger enhancement in the region $m_{jj} < M_W$.

\begin{figure}[ht]
\begin{center}
\includegraphics[angle=0,width=7.5cm]{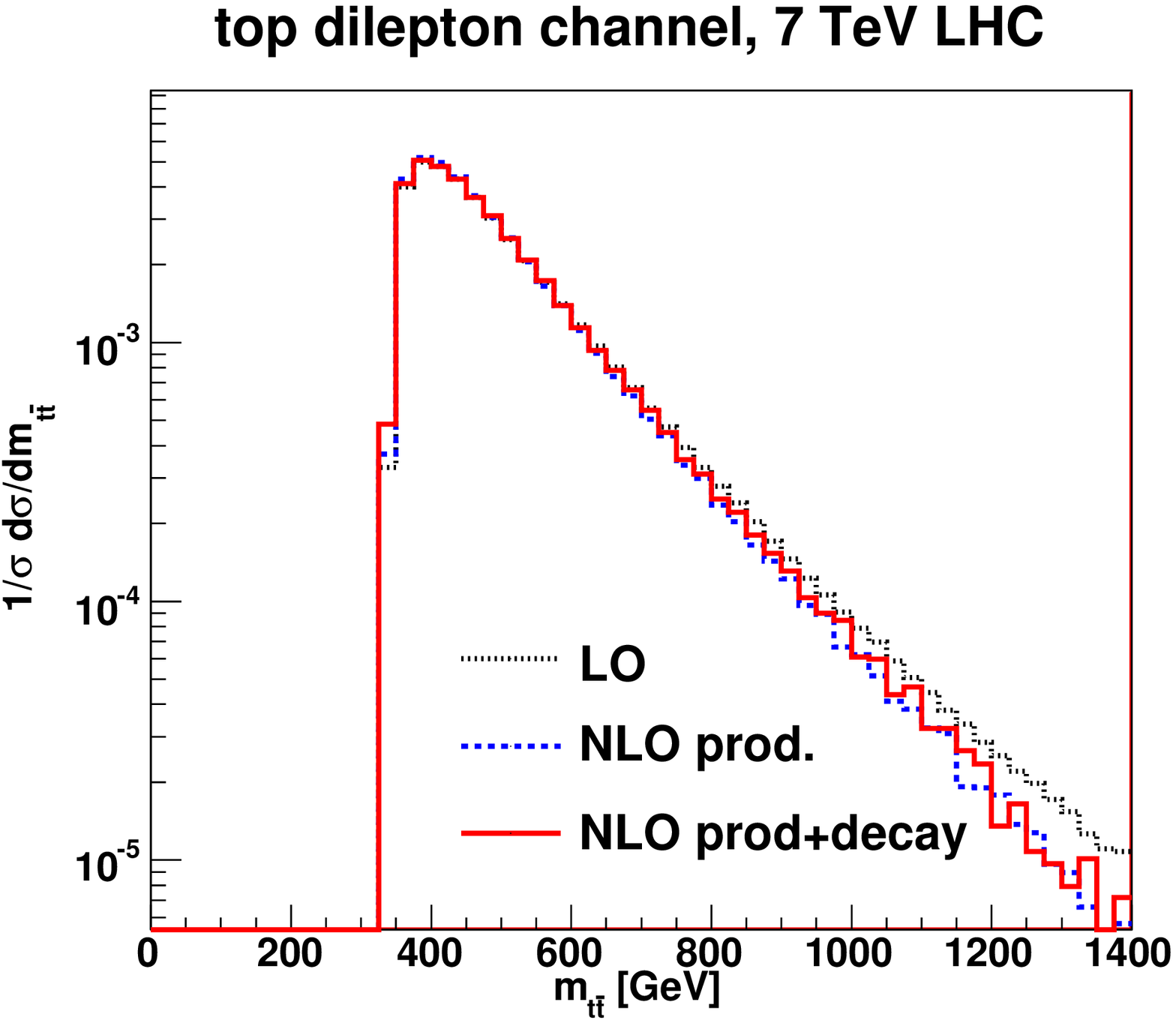}
\includegraphics[angle=0,width=7.5cm]{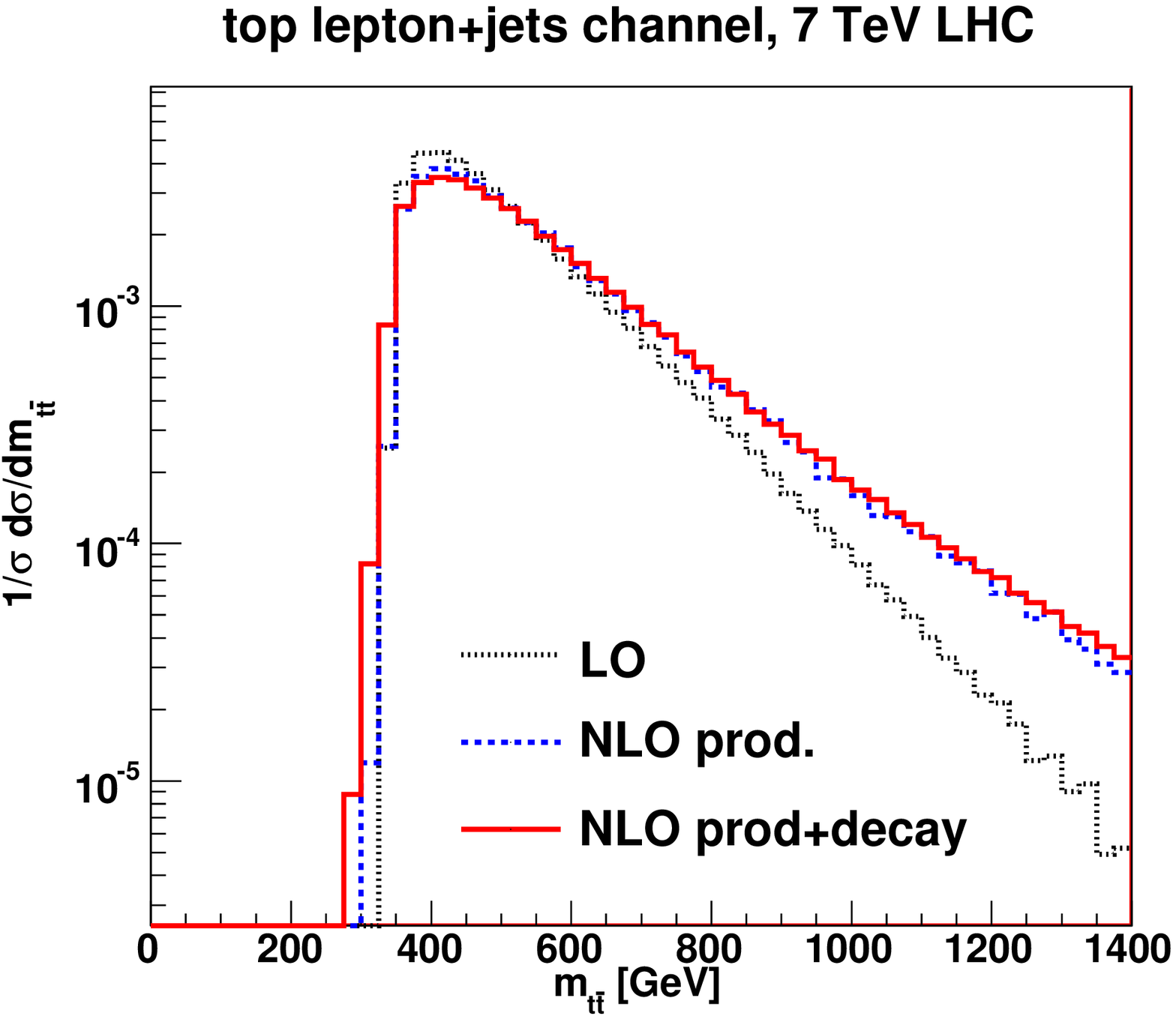}
\caption{Distribution of the invariant mass of the top, anti-top quark
system at the 7 TeV LHC, for the dilepton channel (left) and the
lepton+jets channel (right).}
\label{fig:pasljmttbsyst}
\end{center}
\end{figure}

In Figure~\ref{fig:pasljmttbsyst} we show the invariant mass of the top,
anti-top system in both the dilepton and lepton+jets channels. This distribution
is important for New Physics searches and, in both cases, although we observe
important NLO effects at high $m_{t\bar t}$, this is purely due to the treatment
of NLO effects in the production stage and not in the decay.

\begin{figure}
\begin{center}
\includegraphics[angle=0,width=7.5cm]{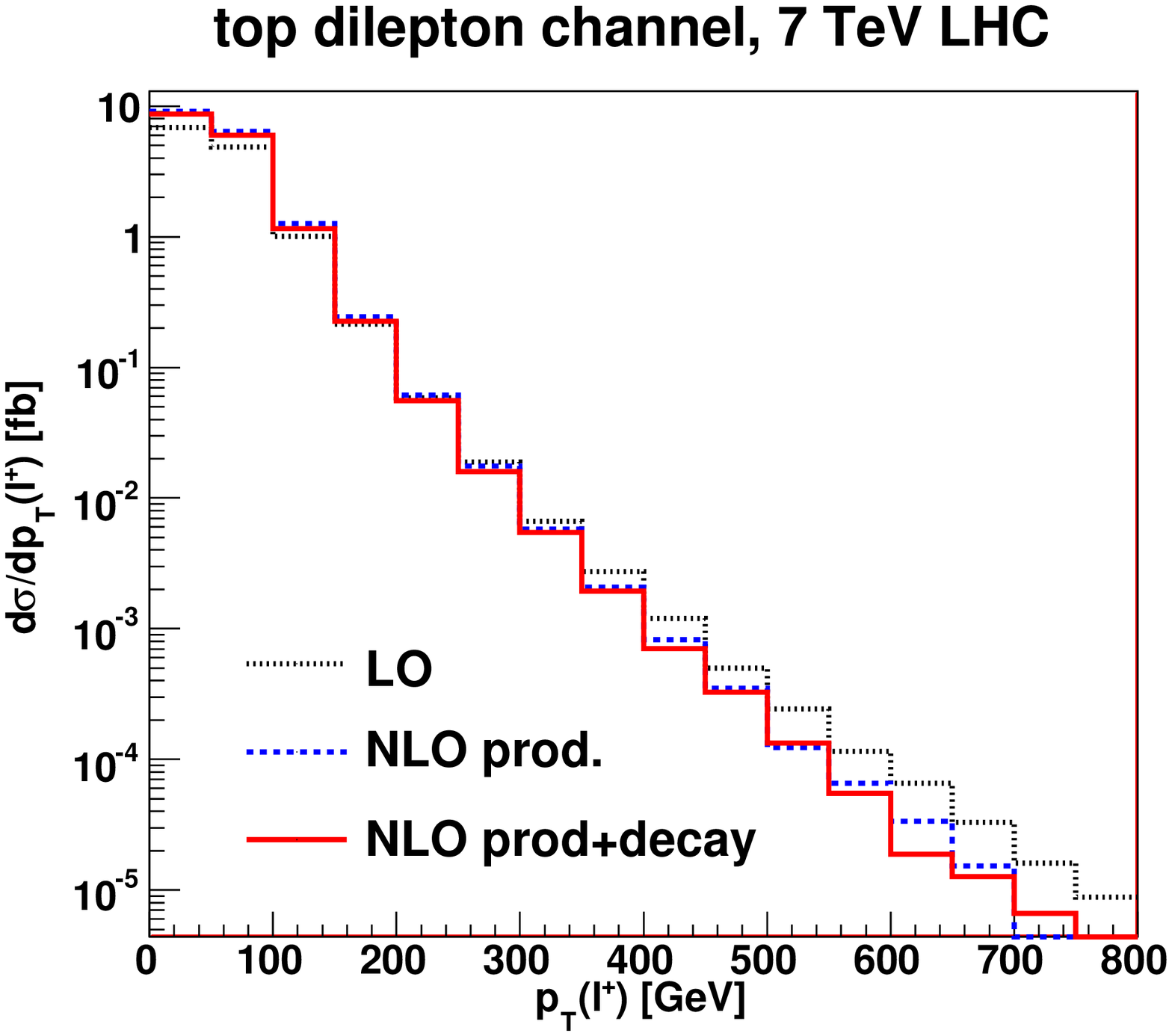}
\includegraphics[angle=0,width=7.5cm]{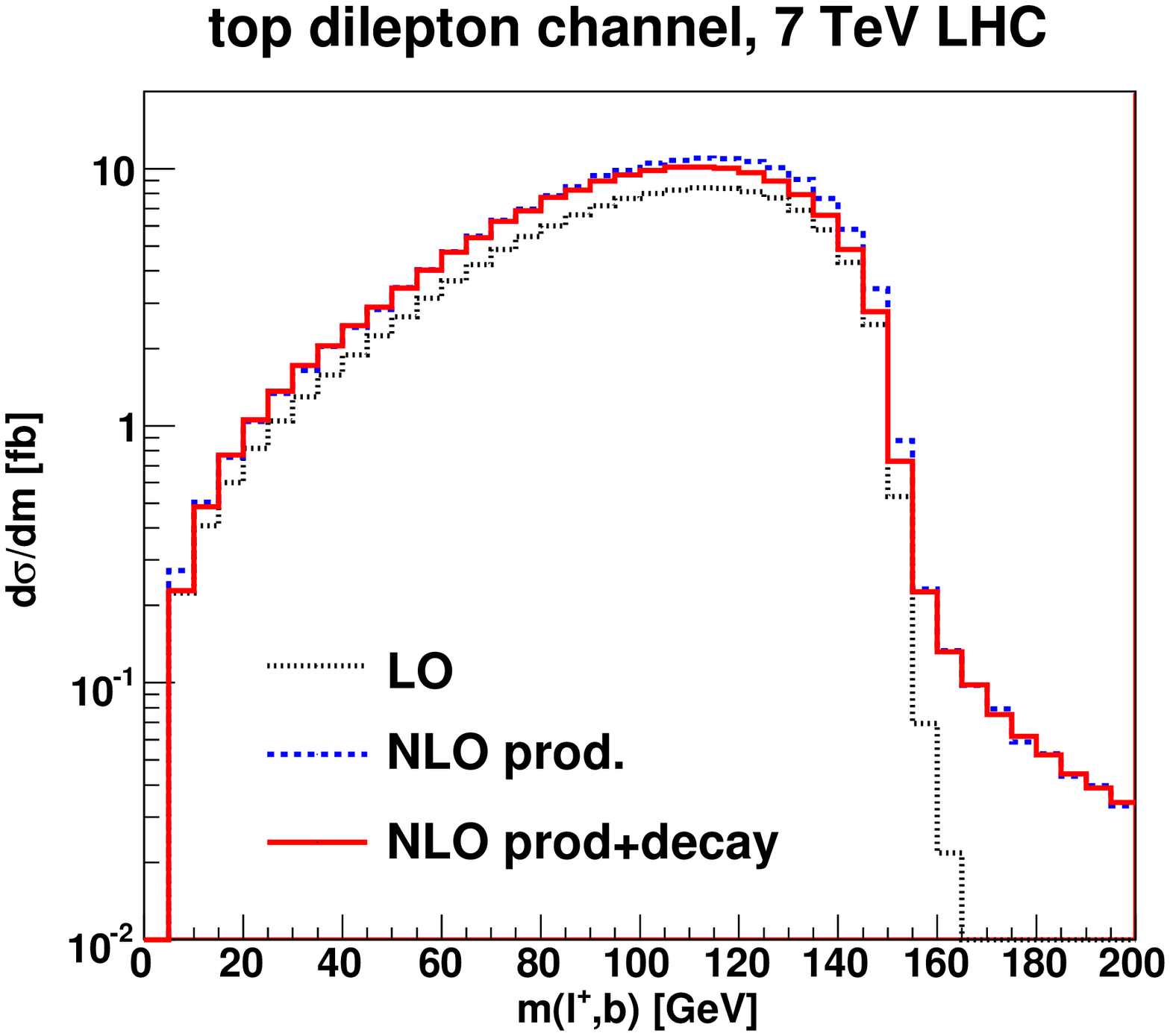}
\caption{Distribution of the transverse momentum of the positively charged lepton, $\ell^+$ (left)
and the invariant mass of the $\ell^+$ and $b$-quark, $m_{\ell^+ b}$ (right) in the top
pair dilepton channel at the 7 TeV LHC.}
\label{fig:kirillcomp}
\end{center}
\end{figure}

Finally in Figure~\ref{fig:kirillcomp} we present the distributions of two additional
observables for the dilepton channel - the transverse momentum of the positively
charged lepton, $p_T(\ell^+)$ and the invariant mass of the $\ell^+$ and $b$-quark
system, $m_{\ell^+,b}$. As first pointed out in Ref.~\cite{Melnikov:2009dn},
in the context of the LHC operating at 10 TeV, the latter distribution is particularly
interesting since it exhibits differences between the NLO predictions
with and without including QCD corrections in the decay (in the region just below the
LO threshold at $m_t$). Here we simply note that the same pattern is observed at 7 TeV, under
typical analysis cuts used for the 2011 run, as was noted in Ref.~\cite{Melnikov:2009dn}.
In passing we also remark that our predictions are in complete agreement with those
of Ref.~\cite{Melnikov:2009dn} once the appropriate differences in input parameters
are accounted for.

\subsubsection{The top quark forward-backward asymmetry at the Tevatron}

A feature of top quark production at the Tevatron that has recently received much attention
is the top quark forward-backward asymmetry~\cite{Aaltonen:2011kc,CDFNote10436,Abazov:2011rq,CDFNote10807}.
The predicted asymmetry is only non-zero at NLO and beyond and, as an example of the utility of our
calculation, in this section we provide a prediction for the parton-level asymmetry within the fiducial
coverage of the CDF detector.

To that end, we focus on the lepton+jets channel and adapt an analysis approach based on that presented
in a recent CDF note~\cite{CDFNote10807}. Partons are clustered using the anti-$k_T$ algorithm 
with distance parameter $D=0.4$ and we require that at least four jets are found that satisfy,
\beq
p_T(\mbox{jet}) > 20~\mbox{GeV} \;, \qquad y(\mbox{jet}) < 2 \;.
\eeq
In addition, we demand that both the $b$ and $\bar b$ quarks are identified within these
jets, with those jets detected in the restricted rapidity range, $y(b-\mbox{jet}) < 1$.
For the leptonic top decay we require,
\beq
p_T(\mbox{lepton}) > 20~\mbox{GeV} \;, \qquad y(\mbox{lepton}) < 1.5 \;,
\eeq
and at least $20$~GeV of missing transverse momentum (from the neutrino).
The top and anti-top quarks are reconstructed using the same procedure as in the LHC lepton+jets analysis
described above.

As for all the previous results, our predictions are based on the choice of scale $\mu_R = \mu_F = m_t$.
Ordinarily a NLO prediction is relatively insensitive to the choice of scale but, in this
case, since the asymmetry is absent in the LO prediction our results depend rather strongly on the
choice of scale. For this reason we also consider variations of this scale by a factor of two about this
central choice, i.e. in the range $(m_t/2, 2m_t)$, and take the variation in the prediction as an
estimate of the theoretical uncertainty due to uncalculated higher orders. 
For each calculation we compute the asymmetry,
\beq
A_{\rm FB} = \frac{\sigma_{\rm NLO}\left(\Delta y > 0\right) - \sigma_{\rm NLO}\left(\Delta y < 0\right)}
 {\sigma_{\rm NLO}\left(\Delta y > 0\right) + \sigma_{\rm NLO}\left(\Delta y < 0\right)} \;,
\label{eq:AFBdefn}
\eeq
where the rapidity difference is defined by $\Delta y = y_t - y_{\bar t}$.
Our results are presented in Table~\ref{table:Afb} where, for the sake of comparison, we also include
predictions for the asymmetry in the absence of any cuts on the decay products of the top quarks.
In addition to computing $A_{\rm FB}$ inclusively, we also present predictions broken down
into contributions over various rapidity difference ranges.
As can be seen from Table~\ref{table:Afb} the theoretical uncertainties permit excursions from the
central value of as much as 60\%. Had we used the leading order prediction in the denominator of Eq.~(\ref{eq:AFBdefn})
the estimated theoretical uncertainties would have been smaller. As such, the use of the NLO result in the
denominator is a conservative choice.
We observe that the asymmetry expected in the fiducial range is smaller than that predicted in
the full rapidity range, but grows with $\Delta y$. The predictions for the two NLO calculations, with
and without NLO effects in the decay, do not differ greatly. Since the component of the calculation that
includes radiation in the decay contains no asymmetry one might expect the computations that include these
effects to result in a smaller value of $A_{\rm FB}$. In contrast, we find that in the region $\Delta y > 0.5$
the cross section under the experimental cuts is lower in the presence of radiation in the decay, so that the denominator in
Eq.~(\ref{eq:AFBdefn}) is reduced and the predicted $A_{\rm FB}$ higher.
\renewcommand{\baselinestretch}{1.6}
\begin{table}
\begin{center}
\begin{tabular}{|rcl|c|c|c|}
\hline
\multicolumn{3}{|c|}{Rapidity range} & $A_{\rm FB}^{\rm no~cuts}$(NLO) 
& $A_{\rm FB}^{\rm cuts}$(NLO production) &  $A_{\rm FB}^{\rm cuts}$(NLO prod + decay) \\
\hline
\multicolumn{3}{|c|}{inclusive} & $0.065^{+0.028}_{-0.014}$  & $0.044^{+0.017}_{-0.010}$ & $0.045^{+0.021}_{-0.011}$ \\
$0 < $&$|\Delta y|$&$ < 0.5$    & $0.036^{+0.020}_{-0.003}$  & $0.017^{+0.005}_{-0.005}$ & $0.015^{+0.009}_{-0.004}$ \\
$0.5 < $&$|\Delta y|$&$ < 1$    & $0.062^{+0.013}_{-0.028}$  & $0.052^{+0.019}_{-0.012}$ & $0.053^{+0.025}_{-0.011}$ \\
$1 < $&$|\Delta y|$&$ < 1.5$    & $0.101 ^{+0.060}_{-0.006}$ & $0.086^{+0.035}_{-0.017}$ & $0.092^{+0.039}_{-0.021}$ \\
& $|\Delta y|$&$ > 1.5$         & $0.193 ^{+0.058}_{-0.036}$ & $0.142^{+0.059}_{-0.034}$ & $0.149^{+0.062}_{-0.036}$ \\
\hline
\end{tabular}
\renewcommand{\baselinestretch}{1}
\caption{Predictions for the top forward-backward asymmetry in the
lepton+jets channel at the Tevatron, computed without applying any experimental
cuts ($A_{\rm FB}^{\rm no~cuts}$) and also when using the cuts described
in the text ($A_{\rm FB}^{\rm cuts}$). The uncertainties are obtained by varying the scale in the range $(m_t/2, 2m_t)$.} 
\label{table:Afb}
\end{center}
\end{table}
\renewcommand{\baselinestretch}{1}

\begin{figure}
\begin{center}
\includegraphics[angle=0,width=7.5cm]{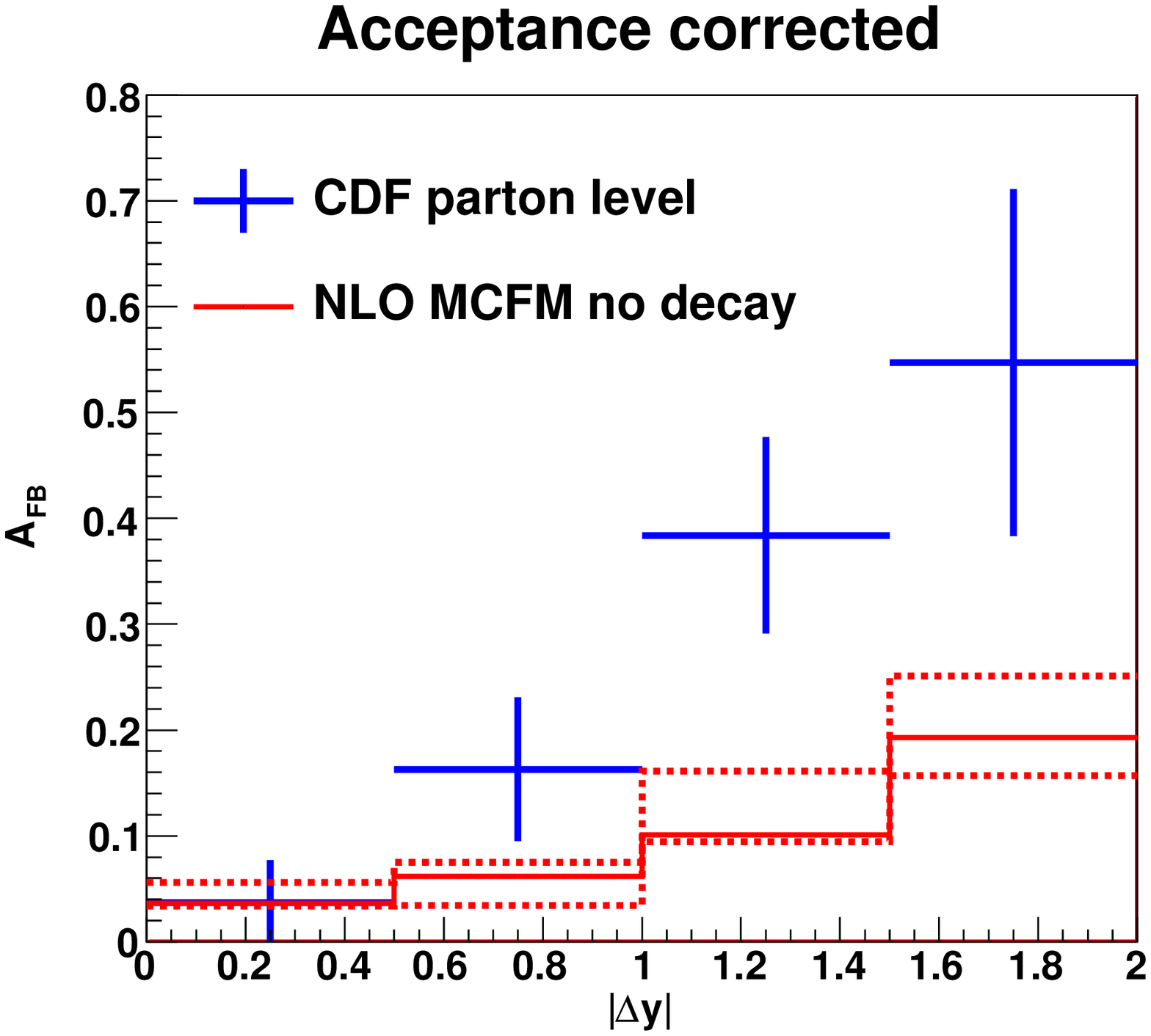}
\includegraphics[angle=0,width=7.5cm]{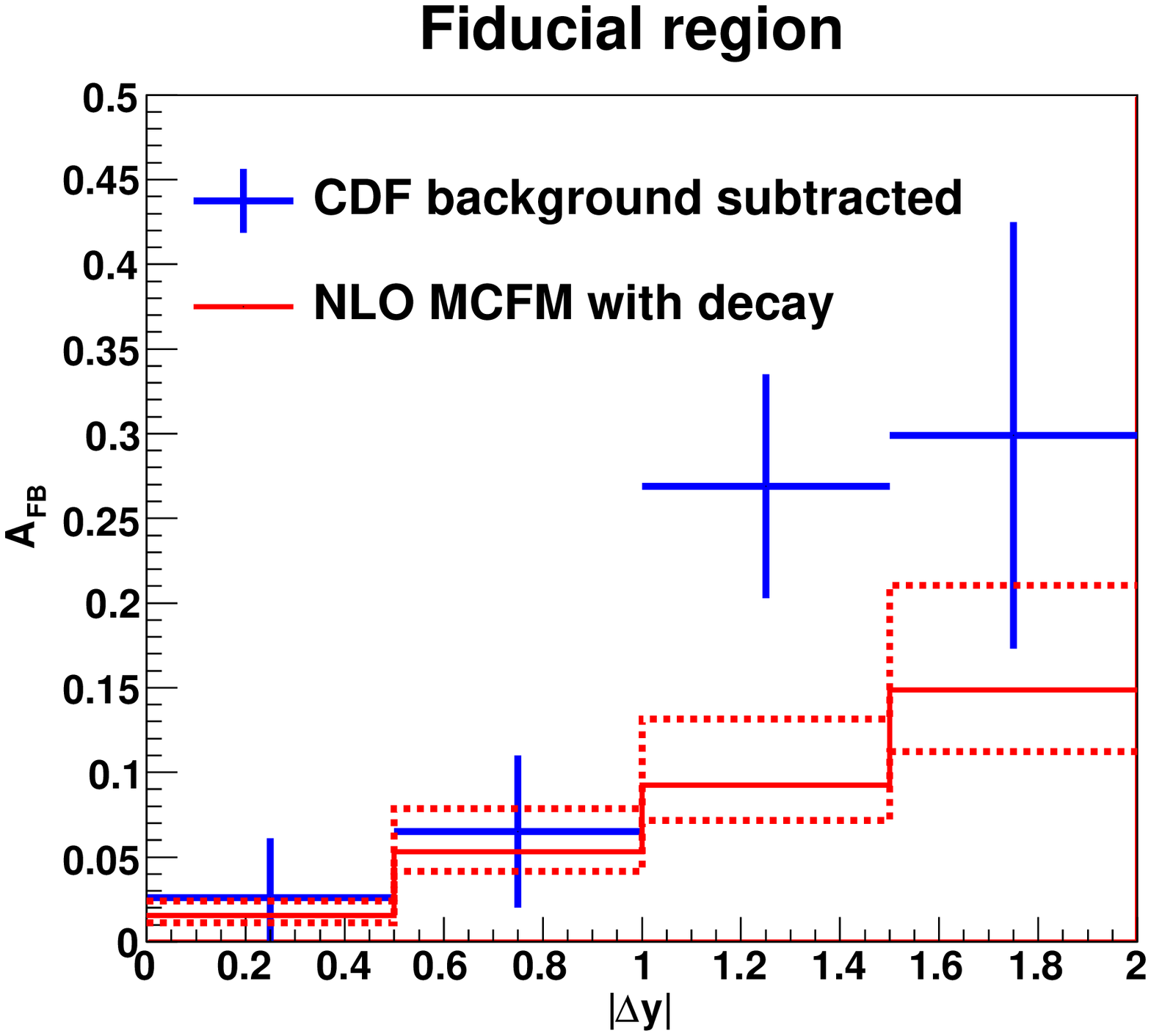}
\caption{The top quark forward-backward asymmetry, $A_{\rm FB}$, at the
Tevatron, after acceptance for corrections (left) and in the measured fiducial region
as detailed in the text (right). Data are taken from CDF~\cite{CDFNote10807} (Tables IX and XVI) and
the MCFM predictions are for inclusive $t{\bar t}$ production (left) and for
the differential rate including radiation in both production and decay (right).
The central prediction is shown as a solid line and the dashed lines represent an estimate of the theoretical uncertainty.}
\label{fig:afb}
\end{center}
\end{figure}

Our results are compared to the CDF results reported in Ref.~\cite{CDFNote10807} in Figure~\ref{fig:afb}.\footnote{
The theoretical predictions in Ref.~\cite{CDFNote10807} include a $26$\% correction to the asymmetry because of electroweak contributions. These corrections will
not be included in our results.}
The inclusive calculation, with no cuts, can be compared in a straightforward manner with the
``parton-level'' corrected results that are presented in the CDF note. In contrast, when the
experimental cuts are applied, the comparison must be interpreted with considerable caution.
Our analysis differs somewhat from that used by CDF, notably in the treatment of the top quark
reconstruction, but most importantly the CDF data is simply background-subtracted and not corrected
back to the parton level. In that case, our theoretical predictions should be interfaced with a
parton shower in order that a full detector simulation be performed before making a definitive comparison.
Nevertheless, our results can be taken
as a guide to the level of uncertainty expected in current predictions of the asymmetry as a function
of $\Delta y$.

We conclude with predictions for an observable that does not depend on reconstructing the top quark
kinematics and that would not be expected to receive significant corrections due to the addition
of a parton shower. This observable is the asymmetry observed in the charge-weighted rapidity distribution
of the lepton (i.e. $q_\ell \eta_\ell$).
The lepton inherits the asymmetry of the top quark and its properties should be predicted more
robustly by our calculation. Using the same procedure for assigning the uncertainty as before we find,
\beq
A_{\rm FB}^{\rm lepton}(\mbox{NLO production}) = 0.021^{+0.006}_{-0.005}\;, \qquad
A_{\rm FB}^{\rm lepton}(\mbox{NLO prod+decay}) = 0.020^{+0.010}_{-0.003}\;.
\eeq
These predictions should be compared with the most recent CDF result~\cite{CDFNote10807},
\beq
A_{\rm FB}^{\rm lepton}(\mbox{CDF}) = 0.065 \pm 0.020 \;.
\eeq

\section{Conclusions}

In this paper we have described in detail the implementation of single
top and top pair processes in the parton-level integrator MCFM. We
have presented an amplitude-level treatment that allows the inclusion
of the decay at NLO for all processes involving top quarks.  We have
currently implemented this scheme in MCFM for three processes,
top-pair production, $s$-channel single top production, and
$t$-channel single top production.

Our treatment of top-pair production has been rewritten to use this
method and to incorporate a faster treatment of the one-loop
amplitudes. 
By retaining the mass for the $b$ quark we can work in a four-flavour
scheme for the $t$-channel single top process. 
In this scheme the
bottom quark does not appear in the initial state, but the production
cross section depends logarithmically on the mass of the $b$-quark.
The four-flavour $t$-channel single top production,
implemented using this method and presented in this paper 
is new. 

The features we have described are complete as of MCFM v6.2.  MCFM
therefore provides the most sophisticated possible NLO treatment of these
top production processes within the context of the top pole
approximation.  With these features we can assess the importance of
NLO radiative effects in production and decay, the effect of the $b$
quark mass, and the effect of an off-shell $W$-boson. For the
distributions that we have examined, these effects turn out to be
quite small.

We look forward to confronting this implementation of NLO theory with data.

\medskip
\noindent
{\bf Acknowledgments} 
We gratefully acknowledge useful conversations with Simon Badger and Markus Schulze.
This research is supported by the US DOE under contract
DE-AC02-06CH11357.
    
\medskip
\appendix

\section{Spinor notation \label{spinnotation}}
Our spinor notation is quite standard in the 
QCD literature, (for a review see refs.~\cite{Mangano:1990by,Dixon:1996wi}). 
The function $u_\pm(k_i)$ is a
massless Weyl spinor of momentum $k_i$ and positive or negative
chirality. In terms of these solutions of the Dirac equation, the spinor products are 
defined by,
\beqn
\spa{i}.{j} &=& 
              \langle i^-|j^+\rangle = \bar{u}_-(k_i)u_+(k_j)\,,
             \label{defspa}\\
\spb{i}.{j} &=& 
              \langle i^+|j^-\rangle = \bar{u}_+(k_i) u_-(k_j)\,.
             \label{defspb}
\eeqn
We use the convention $\spb{i}.{j} = \mathop{\rm
sgn}(k_i^0k_j^0)\spa{j}.{i}^*$, so that,
\beq
\spa{i}.{j}\spb{j}.{i} = 2k_i\cdot k_j \equiv s_{ij} \,.
\eeq
and \beq
\spa{j}.{i} = - \spa{i}.{j} \,, \qquad\quad \spb{j}.{i} = - \spb{i}.{j} \,.
\eeq
For massless spinors we have the following generalization of the Fierz identity,
\beq \label{Fierz}
\langle a| \gamma^\mu | b] \; \gamma_\mu =2 \left( | a \rangle [b|+|b]\langle a| \right) \, .
\eeq
Another useful identity is the Schouten identity,
\beqn
\spa{a}.{b}\spa{c}.{d} &=&
\spa{a}.{d}\spa{c}.{b} + \spa{a}.{c}\spa{b}.{d} \,, \nonumber \\
\spb{a}.{b}\spb{c}.{d} &=&
\spb{a}.{d}\spb{c}.{b} + \spb{a}.{c}\spb{b}.{d} \,.
\label{Schouten}
\eeqn
We further define,
\beq
\langle a|\slsh{p}|b]\ \equiv  \langle a|p|b] \to
\spa{a}.{p}\spb{p}.{b} \,\mbox{for~massless}~p, \qquad\quad
\label{spinorsandwich}
\eeq
where the decomposition into a pair of spinor products is only valid for a
massless momentum $p$.

\section{Calculation of the total width for top decay}
\label{sec:totalwidth}

Our calculation retains the NLO corrections to the fully differential
top decay rate. This requires the inclusion of the correlations with
the decay products of the $W$-boson as well as the effects of extra
gluon radiation if present. The calculation of the total width that we
detail here sums over the polarizations of the $W$-boson and
integrates over the momentum emitted gluon.
Although this is not a new result, 
the total width is one of the ingredients in our calculation.
In addition, the calculation of the one-loop contribution to the
differential rate is almost identical to the calculation of the
one-loop contribution to the total width and our integration over
the real radiation counterterm closely parallels the integration over
the real radiation given below. Our treatment of the corrections to the
width follows, and in places supplements, the nice discussion given in ref.~\cite{Czarnecki:1990kv}.

\subsection{Phase space for tree graph top decay}
In the rest frame of the top it is straightforward to obtain the two
particle phase space  for the decay $t \rightarrow W+ b$,
\beqn
d \Phi^{(2)}(\pt;\pw,\pb) &=&  \frac{d^n \pb}{(2 \pi)^{n-1}}
\frac{d^n \pw}{(2 \pi)^{n-1}} \, (2 \pi)^n \, \delta^n(\pt-\pw-\pb) \; 
\delta(\pb^2-m_b^2) \; 
\delta(\pw^2-m_W^2) \nn \\
&=&\frac{(4 \pi)^{2 \ep}}{32 \pi^2} 
\frac{1}{(m_t^{2})^\ep} \;
\Big[\lambda(1,\omega^2,\beta^2)\Big]^{\frac{1}{2}- \ep} \, d ^{n-2}\Omega_w \;,
\label{eq:PS2}
\eeqn
where $\omega = m_W/m_t$, $\beta = m_b/m_t$ and $\lambda(x,y,z)$ is defined in Table~\ref{Notation}.
The matrix element squared for this process, summed over spins, is,
\beqn
{\cal M}_\mu{\cal M}^*_\nu 
\Bigg( -g^{\mu \nu}+\frac{\pw^\mu \pw^\nu}{m_W^2}\Bigg)
&= & \frac{2 G_F m_t^4}{\sqrt{2}} f \, ,
\eeqn
with $f$ also defined in Table~\ref{Notation}.
Using the fact that,
\beq
\int d^{n-2} \Omega_w = (4\pi)^{1-\ep} \, \frac{\Gamma(1-\ep)}{\Gamma(2-2\ep)} \;,
\eeq
taking the limit $\epsilon \to 0$ and including the flux factor $(2m_t)^{-1}$, we recover the result
for the lowest order width quoted in Eq.~(\ref{eq:LOwidth}).

\subsection{Virtual corrections to top decay}
\label{Virtual_corrections}
\noindent
The general structure of the virtual corrections to the process $t \to W + b$
has been outlined in section~\ref{sec:topdecayvirt}. Here we will give results for
the coefficients $C_0^L$, $C_0^R$, $C_1^L$ and $C_1^R$ that appear in the form
factor decomposition, Eq.~(\ref{eq:massiveF}). 

These coefficients involve the following functions related to the scalar integrals,
\begin{eqnarray}
 A_0(m_1)  &=&
 \frac{\mu^{4-D}}{i (2\pi)^{D}\cg}\int d^D l \;
 \frac{1}{(l^2-m_1^2+i\varepsilon)}\,, \nn \\
 B_0(p_1;m_1,m_2)  &=&
 \frac{\mu^{4-D}}{i (2\pi)^{D}\cg}\int d^D l \;
 \frac{1}
{(l^2-m_1^2+i\varepsilon)
((l+p_1)^2-m_2^2+i\varepsilon)}\,, \\
 C_0(p_1,p_2;m_1,m_2,m_3)  &=&
\frac{\mu^{4-D}}{i (2\pi)^{D}\cg} 
 \int d^D l \;\frac{1}{(l^2-m_1^2+i\varepsilon)((l+p_1)^2-m_2^2+i\varepsilon)((l+p_1+p_2)^2-m_3^2+i\varepsilon)}\,.
 \nn
\eeqn
where the constant $\cg$ has been defined in Eq.~(\ref{cGamma}).
\renewcommand{\baselinestretch}{1.8}
\begin{table}
\begin{center}
{\large 
\begin{tabular}{|l|l|}
\hline
Integral function & Expression \\
\hline
$A_0(m_t) $  &  $ m_t^2 \big(\frac{1}{\epsilon}+\ln\frac{\mu^2}{m_t^2}+1\big) +O(\epsilon)  $ \\
$A_0(m_b) $  &  $ m_b^2 \big(\frac{1}{\epsilon}+\ln\frac{\mu^2}{m_t^2}+1 -\ln \beta^2 \big) +O(\epsilon)  $ \\
$B_0(\pt;0,m_t) $  &  $ \frac{1}{\epsilon}+\ln\frac{\mu^2}{m_t^2}+2 +O(\epsilon)  $ \\
$B_0(\pb;0,m_b) $  &  $ \frac{1}{\epsilon}+\ln\frac{\mu^2}{m_t^2}+2 -\ln(\beta^2)+O(\epsilon)  $ \\
$B_0(\pw;m_b,b_t) $  &  $ \frac{1}{\epsilon}+\ln\frac{\mu^2}{m_t^2}+2 +b^{(f)}+O(\epsilon)  $ \\
$C_0(\pb,\pt;0,m_b,m_t)  $  &  $ \frac{1}{2 m_t^2 \bar{P}_3}\left[
 \bar{Y}_p \Big(\frac{1}{\epsilon}+\ln\frac{\mu^2}{m_t^2}\Big)+c^{(f)}+O(\epsilon)\right] $ \\
\hline
\end{tabular}
\caption{Table of one-loop integrals obtained from ref.~\cite{Ellis:2007qk}. 
The functions $c^{(f)}$ and $b^{(f)}$ are given in the text.}
\label{Oneloop}}
\end{center}
\end{table}
\renewcommand{\baselinestretch}{1}
Explicit expressions for the particular scalar integrals that appear in the calculation 
are given in Table~\ref{Oneloop}.
In order to have a compact representation for the finite
parts of the integrals, we have introduced the following functions,
\beqn
b^{(f)}& = &
\frac{(1-\omega^2-\beta^2)}{\omega^2 } \ln \beta + \frac{2\bar{P}_3}{\omega^2} \, \bar{Y}_p \;,
\nonumber \\
c^{(f)} &=& 
      \li(1-\bar{P}_{-})-\li(1-\bar{P}_{+})-\li\left(1-\frac{\bar{P}_{-}}{\bar{P}_{+}}\right)+\ln^2 \beta- \ln^2 \bar{P}_{+} \;.
\eeqn

The results for the coefficients in the four-dimensional helicity scheme~\cite{Bern:2002zk} are, 
\beqn \label{resultsforCcoefficients}
C_0^L&=&\left[\frac{{\Pzb}}{\P3b} \Ypb -1\right] \frac{2}{\epsilon}+\frac{2 \Pzb}{\P3b} c^{(f)} 
 -4+\ln(\beta^2)
\nonumber  \\
       &+&\frac{1}{2\P3b^2} b^{(f)} \left[1-\beta^2 \om^2-\om^2+\beta^4-2 \beta^2-6 \P3b^2\right]
       -\frac{1}{2\P3b^2} \ln(\beta^2) \left[3 \P3b^2+\beta^2 \om^2-\beta^4+\beta^2\right] \;, \nonumber \\
C_0^R&=&\frac{\beta}{\P3b^2} \left[\om^2 b^{(f)}-\frac{1}{2} (1-\beta^2-\om^2) \ln(\beta^2)\right] \;, \nonumber \\
C_1^L&=&\frac{\beta}{\P3b^2} \left[\Pzb \ln(\beta^2)-\Wzb b^{(f)}\right] \;, \nonumber \\
C_1^R&=&\frac{1}{\P3b^2} \left[\frac{1}{2} (1-\om^2-\beta^2) b^{(f)}-\beta^2 \ln(\beta^2)\right] \;.
\label{C-coefficients}
\eeqn
These expressions include the effects of wave function renormalization, as indicated in the
upper-left diagram of Figure~\ref{tdecayg}.
The result in the four-dimensional helicity scheme is,
\begin{equation} \label{wavefunction}
Z_Q= 1-g^2 \cg C_F \Bigg[ \frac{3}{\ep}+3 \ln\left(\frac{\mu^2}{m^2}\right)+5\Bigg]+O(g^4,\ep)\; .
\end{equation}
The result for the wave function renormalization is independent of 
the gauge fixing parameter.

In terms of these coefficients the matrix element squared is,
\beqn
&&{\cal M}_\mu{\cal M}^*_\nu 
\Bigg[ -g^{\mu \nu}+\frac{\pw^\mu \pw^\nu}{m_W^2}\Bigg]= \nonumber \\
&&\frac{2 G_F m_t^4}{\sqrt{2}} \Bigg\{f + 2 g^2\cg 
\left( \frac{\mu^2}{m_t^2}\right)^\epsilon C_F \Bigg[ 
C_0^L f+(C_1^R+\beta C_1^L) (\frac{f}{2}-3 \om^2 \Pzb)-6 \beta \om^2 C_0^R\Bigg]\Bigg\} +O(\as^2) \;.
\eeqn

Inserting the values for the coefficients from Eq.~(\ref{C-coefficients}) and including the wave-function 
renormalization the total virtual result is, 
\beqn \label{Czarneckivirtresult}
\as \Gamma_1^{\rm virt} &= & 
\Gamma_\infty \, 4 C_F g^2 \cg \left( \frac{\mu^2 }{m_t^2} \right)^{\ep} 
\Bigg[ 2 f \bigg\{
          \frac{1}{\epsilon} \big[\Pzb\Ypb-\P3b]  
          + \Pzb \Big[\li(1-\Pmb)-\li(1-\Ppb)-\li\left(1-\frac{\Pmb}{\Ppb}\right)\nonumber \\
          &+&\Ypb^2 - 2 (\Ypb+ \ln\beta ) (\Ywb+\Ypb)+ 2 \Ywb  \ln \Ppb \Big]-2 \P3b \bigg\} \nonumber \\
       &+& 12 \om^2 \Ypb \P3b^2 
  - \big[1  + 4 \beta^2- 5 \beta^4 - \om^2 (5- \beta^2) +4 \om^4\big] \P3b  \ln\beta \Bigg]\, ,
\eeqn
in agreement with Czarnecki~\cite{Czarnecki:1990kv} after using standard identities between dilogarithms.

\subsection{Real radiation}
The evaluation of the real corrections relies on a factorization of the phase space
into a simple form that is suited both to the calculation of the total width and the
more differential calculations presented in this paper.

\subsubsection{Factorization of Phase Space}
For radiation in the decay the appropriate phase space is given by,
\beqn
d\Phi^{(3)}(p_t; \pw, \pb, \pg)
 &=& \frac{d^n \pw}{(2\pi)^{n-1}} \frac{d^n \pb}{(2\pi)^{n-1}} \frac{d^n \pg}{(2\pi)^{n-1}} \nn \\
 &\times& \delta^+(\pw^2-m_W^2) \delta^+(\pb^2-m_b^2) \delta^+(\pg^2) 
    (2\pi)^n \delta^n(p_t-\pw-\pb-\pg) \;.
\eeqn
By inserting an integral over a delta function, $\delta^n(p_t-\pw-P) d^nP$, we can rewrite this in the
familiar factorized form,
\beq
d\Phi^{(3)}(p_t; \pw, \pb, \pg) =
\int \frac{dP^2}{2\pi} \, d\Phi^{(2)}(p_t; \pw, P) \, d\Phi^{(2)}(P; \pb, \pg) \;,
\label{eq:psfact} 
\eeq
where $P=\pb+\pg$.
Working in the center of mass frame of $P=\pb+\pg$ we have,
\beqn
\pg &=& \frac{m_t (z-\beta^2)}{2 \sqrt{z}}(1,\ldots, \sin\theta, \cos\theta ) \;, \nn \\
\pb &=& \frac{m_t (z+\beta^2)}{2 \sqrt{z}}(1,\ldots, -v \sin\theta, -v \cos\theta ) \;, \quad \mbox{with}~
v   = \frac{z-\beta^2}{z+\beta^2} \;, \nn \\
p_t &=& (\Et,\ldots,0, \Pt ) \;, \quad \mbox{with}~
\Et = \frac{m_t}{2 \sqrt{z}}(1+z-\omega^2),\;\;
\Pt = \frac{m_t}{2 \sqrt{z}} \sqrt{\lambda(1,\omega^2,z)} \;.
\eeqn
Hence $P^2=m_t^2 z$ and we can use the result of Eq.~(\ref{eq:PS2}) to write the factorized
form, Eq.~(\ref{eq:psfact}), as,
\beqn
d\Phi^{(3)}(p_t; \pw, \pb, \pg) &=&
\frac{m_t^2}{2\pi} \int dz \;
\frac{(4 \pi)^{2 \ep}}{32 \pi^2} 
\frac{1}{(m_t^{2})^\ep} \;
\Big[\lambda(1,\omega^2,z)\Big]^{\frac{1}{2}- \ep} \, d ^{n-2}\Omega_w \nn \\
&& \times
\frac{(4 \pi)^{2 \ep}}{32 \pi^2} 
\frac{1}{(m_t^{2})^\ep} \;
z^{-1+\ep} (z-\beta^2)^{1-2\ep} \, d ^{n-2}\Omega_g \;.
\label{eq:PS3}
\eeqn
By comparing Eq.~(\ref{eq:PS3}) with the lowest order
space space in Eq.~(\ref{eq:PS2}) we can
write the real emission phase space as a leading
order normalization multiplying a term containing the real
emission degrees of freedom,
\beq
d\Phi^{(3)} (p_t; \pw, \pb, \pg)
 = d \Phi^{(2)} (\pt; \pw, \pb) \times [dg(\pt ,\pw, z)] \;.
\label{eq:PSfactorized}
\eeq
The real emission factor factor $[dg(\pt ,\pw, z)]$ is given by,
\beqn
[dg(\pt ,\pw, z)] &=&
\frac{(m_t^2)^{1-\ep} (4 \pi)^{2 \ep}}{64 \pi^3} \int dz 
\Big[\frac{\lambda(1,\omega^2,z)}
 {\lambda(1,\omega^2,\beta^2)}\Big]^{\frac{1}{2}- \ep}  
z^{-1+\ep} (z-\beta^2)^{1-2\ep} \, d ^{n-2}\Omega_g \nn \\
&=& 
c_\Gamma (m_t^2)^{1-\ep} \int dz 
\Big[\frac{\lambda(1,\omega^2,z)}
 {\lambda(1,\omega^2,\beta^2)}\Big]^{\frac{1}{2}- \ep}  
z^{-1+\ep} (z-\beta^2)^{1-2\ep}
 \left[ \frac{4^{\ep}}{2} \int_{-1}^1 dx (1-x^2)^{-\ep} \right] \;.
\label{eq:finalfact}
\eeqn
In this equation  we have set $x=\cos \theta$ and $\cg$ has been defined
in Eq.~(\ref{cGamma}).

\subsubsection{Matrix element squared and integration.}
We now present the result for the spin-averaged matrix element squared
for the real radiation process, $t \to W + b + g$. The result is,
\beqn
{\cal M}_\mu{\cal M}^*_\nu 
\Big( -g^{\mu \nu}+\frac{\pw^\mu \pw^\nu}{m_W^2}\Big) &=&
  2 \gs^2 C_F \frac{G_F m_t^4}{\sqrt{2}} \Bigg\{ 
  f \left[-\frac{m_t^2}{p_t.p_g^2}-\frac{m_b^2}{p_b.p_g^2}
         +\frac{2 m_t^2 \Pzb}{ p_t.p_g \, p_b.p_g} \right] \label{eq:realradmsq} \\
&+&
  f \left[\frac{2 p_b.p_g-2 p_t.p_g}{ p_t.p_g \, p_b.p_g} \right]
 +\frac{4 \om^2}{m_t^2} \frac{(p_t.p_g^2+p_b.p_g^2)}{p_t.p_g \, p_b.p_g}
  +\frac{2(\beta^2+1)}{m_t^2} \frac{(p_t.p_g-p_b.p_g)^2}{p_t.p_g \, p_b.p_g} \Bigg\} \;. \nn
\eeqn
This expression can be written in terms of the variables of integration in the
phase space measure, $z$ and $x=\cos\theta$ (c.f. Eq~(\ref{eq:finalfact})), by using the
identities,
\beq
p_b.p_g = \frac{m_t^2}{2} (z-\beta^2) \; \qquad p_t.p_g = E_g (E_t - P_t\cos\theta) \;.
\eeq
It is convenient to first perform the angular integration over $x =\cos \theta$. From
the expression for the matrix element in Eq.~(\ref{eq:realradmsq}) and the form of the phase space
in Eq.~(\ref{eq:finalfact}), these integrations
take the form,
\beq
J_n = \frac{4^\ep}{2} \int_{-1}^{1} dx \; (1-x^2)^{-\ep} \frac{1}{(p_t \cdot p_g )^n}
\label{eq:Jdefn}
\eeq
for $n=0$,~$1$ and $2$. The results for $J_0,J_1$ and $J_2$ are given in Table~\ref{angint}.
\renewcommand{\baselinestretch}{1.8}
\begin{table}
\begin{center}
{\large 
\begin{tabular}{|l|l|}
\hline
Integral & Result \\
\hline
$J_0$
& $ 1 + 2 \epsilon + O(\epsilon^2)$ \\
$J_1$
& $\frac{1}{E_g \Pt} \Big\{Y_p+\epsilon 
\left[ \li\left(1-\frac{P_{-}}{P_{+}}\right)+Y_p^2 \right]+O(\ep^2)\Big\}$ \\
$J_2$
& $\frac{1}{E_g^2 m_t^2} \Big\{1 +2 \ep \frac{\Et}{\Pt}Y_p +O(\ep^2)\Big\} $ \\
\hline
\end{tabular}
\caption{Results for the angular integrations specified in Eq.~(\ref{eq:Jdefn}).}
\label{angint}
}
\end{center}
\end{table}
\renewcommand{\baselinestretch}{1}

We shall first focus on the integration of the divergent terms
appearing in the first set of parentheses $[ \ldots ]$ in Eq.~(\ref{eq:realradmsq}).
After performing the integration over $x$ these terms lead to the following
expressions, with the integral over $z$ not yet evaluated.
For the first term, containing the factor $m_t^2/(p_t.p_g)^2$, 
we are left with,
\beqn
S_1
&=&g^2 \cg C_F\; \left(\frac{\mu^2}{m_t^2}\right)^\ep 
\int_{\beta^2}^{z_m} \; dz \; z^{-\ep}
\Big(\frac{z-\beta^2}{z}\Big)^{1-2\ep}
\Big(\frac{\lambda(1,\om^2,z)}{\lambda(1,\om^2,\beta^2)}\Big)^{\frac{1}{2}-\ep} \frac{m_t^2}{E_g^2} 
\Big\{1 +2 \ep \frac{\Et}{\Pt}Y_p +O(\ep^2)\Big\} \nonumber  \\
&=&g^2 \cg  C_F\; \left(\frac{\mu^2}{m_t^2}\right)^\ep 
\int_{\beta^2}^{z_m} \; dz \; 
\frac{4}{(z-\beta^2)^{1+2 \ep}} z^\ep
\Big(\frac{\lambda(1,\om^2,z)}{\lambda(1,\om^2,\beta^2)}\Big)^{\frac{1}{2}-\ep} 
\Big\{1 +2 \ep \frac{\Et}{\Pt}Y_p +O(\ep^2)\Big\} \, .
\label{eq:S1int}
\eeqn
The expression for the second term, with the factor $m_b^2/(p_b.p_g)^2$, is,
\beqn
S_2
&=& g^2 \cg C_F\; \left(\frac{\mu^2}{m_t^2}\right)^\ep 
\int_{\beta^2}^{z_m} \; dz \; z^{-\ep}
\Big(\frac{z-\beta^2}{z}\Big)^{1-2\ep} \frac{4 \beta^2}{(z-\beta^2)^2}
\Big(\frac{\lambda(1,\om^2,z)}{\lambda(1,\om^2,\beta^2)}\Big)^{\frac{1}{2}-\ep} 
\Big\{1 +2 \ep+O(\ep^2)\Big\} \nn \\
&=& g^2 \cg  C_F\; \left(\frac{\mu^2}{m_t^2}\right)^\ep 
\int_{\beta^2}^{z_m} \; dz \; 
\frac{4 \beta^2}{(z-\beta^2)^{1+2 \ep}} z^{-1+\ep}
\Big(\frac{\lambda(1,\om^2,z)}{\lambda(1,\om^2,\beta^2)}\Big)^{\frac{1}{2}-\ep} 
\Big\{1 +2 \ep +O(\ep^2)\Big\} \, .
\label{eq:S2int}
\eeqn
The corresponding result for the third and final term, proportional to $\bar{P}_0/(p_t.p_g \; p_b.p_g)$, is,
\beqn
S_3
&=& -g^2 \cg  C_F\; \left(\frac{\mu^2}{m_t^2}\right)^\ep 
\int_{\beta^2}^{z_m} \; dz \; z^{-\ep}
\Big(\frac{z-\beta^2}{z}\Big)^{1-2\ep} \frac{2 (1+\beta^2-\om^2)}{(z-\beta^2)}
\Big(\frac{\lambda(1,\om^2,z)}{\lambda(1,\om^2,\beta^2)}\Big)^{\frac{1}{2}-\ep} 
\frac{m_t^2}{E_g \Pt} \nn \\
&\times& \Big\{Y_p+\epsilon
\left[\li\left(1-\frac{P_{-}}{P_{+}}\right)+Y_p^2\right]+O(\ep^2)\Big\}\nonumber \\
&=& -g^2 \cg  C_F\; \left(\frac{\mu^2}{m_t^2}\right)^\ep \frac{8 (1+\beta^2-\om^2)}{\lambda(1,\beta,\om^2)}
\int_{\beta^2}^{z_m} \; dz \; z^{\ep}
 (z-\beta^2)^{-1-2\ep} \Big(\frac{\lambda(1,\om^2,z)}{\lambda(1,\om^2,\beta^2)}\Big)^{-\ep} \nn \\
&\times& \Big\{Y_p+\epsilon
\left[\li\left(1-\frac{P_{-}}{P_{+}}\right)+Y_p^2\right]+O(\ep^2)\Big\} \, .
\label{eq:S3int}
\eeqn
We may now perform the integration over $z$ for each of these terms.  
For convenience we give useful results for the non-trivial integrands in Table~\ref{zint}.
The notation is defined by,
\beq \label{fdef}
\langle f(z) \rangle \equiv \int_{\beta^2}^{z_m} \, dz \, f(z) \;.
\eeq
\renewcommand{\baselinestretch}{1.8}
\begin{table}
\begin{center}
{\large 
\begin{tabular}{|l|l|}
\hline
$f(z)$ & $\langle f(z)\rangle$ \\* 
\hline
$\frac{1}{z-\beta^2} \; \left[ Y_p(z)-\Ypb \right]$ &
$\bar{Y}_p \left(\ln(4 \bar{P}_3^2) - \bar{Y}_p \right)
      +\frac{1}{2} \li(1-\bar{P}_{-}) - \frac{1}{2} \li(1-\bar{P}_{+})$ \\* 
      & $ - \li(1-\frac{\bar{P}_{-}}{\bar{P}_{+}}) 
        +\ln(\bar{P}_{-}) \ln(1-\omega-\bar{P}_{-})
        -\ln(\bar{P}_{+}) \ln(\bar{P}_{+}-1+\omega)$ \\* 
$ \frac{1}{z-\beta^2} \left[P_3(z)-\P3b\right]$ &
$\P3b \ln \left(\frac{4 \P3b^2}{\omega (z_m-\beta^2)}\right)-\P3b-\Wzb\Ywb$ \\* 
$  z \Ypz$  & $\frac{1}{4} \P3b (1+\beta^2+5 \om^2)
      -\frac{1}{2} \om^2 (2+\om^2) \Ywb
      -\frac{1}{2} \beta^4 \Ypb $ \\*
$\Ypz $ & $ \P3b-\om^2 \Ywb-\beta^2 \Ypb$ \\*
$ \frac{P_3(z)}{z^2}$ & $ -\frac{(1+\om^2)}{1-\om^2} \Ypb-\frac{\om^2}{1-\om^2} \Ywb+\frac{\P3b}{\beta^2} $ \\*
$\frac{P_3(z)}{z}$ & $(1-\om^2)\Ypb-\om^2 \Ywb-\P3b $ \\*
$P_3(z)$  & $-\om^2 \Ywb+\Wzb \P3b$ \\*
\hline
\end{tabular}
\caption{Results for finite $z$ integrals as defined in Eq.~(\ref{fdef}). Notation as in Table~\ref{Notation}.}
\label{zint}
}
\end{center}
\end{table}
\renewcommand{\baselinestretch}{1}
Performing the integrals in Eqs.~(\ref{eq:S1int}),~(\ref{eq:S2int}) and (\ref{eq:S3int}),
using the results of Table~\ref{zint}, we obtain the following explicit results,
\beqn
S_1&=& -4 g^2 \cg C_F\; \left(\frac{\mu^2}{m_t^2}\right)^\ep
\Bigg\{ -\frac{1}{2 \ep}-1+\ln\left(\frac{4 \Pthreeb^2}{\beta \om}\right)
     - \bar{P}_0  \frac{\Ypb}{\Pthreeb} - \Wzb \frac{\Ywb} {\Pthreeb} \Bigg\} \;,
\label{eq:S1res} \\
S_2&=& -4 g^2 \cg C_F\; \left(\frac{\mu^2}{m_t^2}\right)^\ep 
\Bigg\{ -\frac{1}{2\ep}-1+\ln\left(\frac{4 \Pthreeb^2}{\beta \om}\right)
-(1-\om^2) \frac{\Ypb}{\Pthreeb} -\frac{1}{2} (1-\om^2-\beta^2) \frac{\Ywb}{\Pthreeb} \Bigg\} \;,
\label{eq:S2res} \\
S_3&=&  4 g^2 \cg C_F\; \left(\frac{\mu^2}{m_t^2}\right)^\ep \frac{\bar{P}_0}{\Pthreeb} \Bigg\{ 2 \Ypb 
    \Big[-\frac{1}{2 \ep}+\ln\Big(\frac{4 \Pthreeb^2 (z_m-\beta^2)}{\beta}\Big)\Big]
      +\li(1-\Pmb)-\li(1-\Ppb)\nn \\
&-&3 \, \li\Big(1-\frac{\Pmb}{\Ppb}\Big) - 3 \, \Ypb^2 +2 \ln(\Pmb) \ln(1-\om-\Pmb)-2 \ln(\Ppb) \ln(\Ppb-1+\om) \Bigg\} \;.
\label{eq:S3res}
\eeqn
The remaining finite contributions can be integrated in four dimensions and added to the
above results to yield the full real contribution to the width,
\beqn
\alpha_s \Gamma_1^{\rm real}   &=& 
\Gamma_\infty \, 4 C_F g^2 \cg \left( \frac{\mu^2 }{m_t^2} \right)^{\ep} 
\Big[
       2 f \Big\{
          -  \frac{1}{\epsilon} [\Pzb \Ypb -\P3b] 
  +\Pzb \Big[ \li(1-\Pmb) - \li(1-\Ppb)- 3 \,\li(1-\frac{\Pmb}{\Ppb})\nonumber \\
          &-& \Ypb^2 + 2 (\Ypb+\ln(\beta)) (\Ywb+\Ypb)
          + 2 \ln\left(\frac{4 \P3b^2}{\Wpb \Ppb^2}\right) \Ypb
          \Big]
    +  \Big[ 2 + 2 \ln\left(\frac{\om}{4 \P3b^2}\right)
          \Big]\P3b \Big\} \nonumber \\
       &+& 2  (1-\beta^2)  \big[
           (1-\beta^2)^2
          +  \om^2 (1+\beta^2)
          - 4 \om^4 
          \big] \Ywb
\nonumber \\
       &+&\frac{1}{2} \big[
           3- \beta^2+ 11 \beta^4 - \beta^6- 4 \beta^4 \om^2
          - 9 \om^4+ 7 \beta^2 \om^4+ 6 \om^6 \big] \Ypb
\nonumber \\
       &+&4   \big[1 - 2 \beta^2+ \beta^4 + \om^2+  \beta^2 \om^2 - 2 \om^4 \big] \P3b \ln(\beta) 
\nonumber \\
       &+&\frac{1}{2} \big[
           5- 22 \beta^2+ 5 \beta^4+ 9 \om^2 (1+\beta^2)- 6 \om^4 \big]\P3b \big] \, .
\label{realcorrection}
\eeqn
Adding in the virtual correction as given in Eq.~(\ref{Czarneckivirtresult}) 
we find that the total correction to the width is in agreement with Eq.~(\ref{totalwidth}).

\bibliography{topnlo}
\bibliographystyle{JHEP}

\end{document}